\documentclass[twocolumn]{aastex6}
\pdfoutput=1 

\newcommand {\kss} {km~s$^{-1}$}

\newcommand{\OI}{[O{\sc i}]}

\newcommand{\OIII}{[O\,{\sc iii}]}
\newcommand{\SII}{[S\,{\sc ii}]}
\newcommand{\NII}{[N\,{\sc ii}]}
\newcommand{\NI}{[N\,{\sc i}]}
\newcommand{\OII}{[O\,{\sc ii}]}

\newcommand{\Ha}{H$\alpha\,$}
\newcommand{\Hb}{H$\beta\,$}

\newcommand{\HII}{H\,{\sc ii}}

\shorttitle{A light echo between galaxies}
\shortauthors{Merluzzi et al.}

\begin{document}

\title{An interacting galaxy pair at the origin of a {\it light echo}}


\author{Paola Merluzzi\altaffilmark{1} and Giovanni Busarello}
\affil{INAF-Osservatorio Astronomico di Capodimonte, Salita Moiariello
  16 I-80131 Napoli, Italy}

\author{Michael A. Dopita and Adam D. Thomas} \affil{Research School of
  Astronomy and Astrophysics, Australian National University,
  Canberra, ACT 2611, Australia}

\author{Chris P. Haines} \affil{INAF-Osservatorio Astronomico di
  Brera, Via Brera 28 I-20121 Milano, Italy}

\and

\author{Aniello Grado, Luca Limatola and Amata Mercurio}
\affil{INAF-Osservatorio Astronomico di Capodimonte, Salita Moiariello
  16 I-80131 Napoli, Italy}

\altaffiltext{1}{merluzzi@na.astro.it}


\begin{abstract} 
   In a low-density region of the Shapley supercluster we identified
   an interacting galaxy pair at redshift $z = 0.04865$ in which the
   Seyfert-2 nucleus of the main galaxy (ShaSS\,073) is exciting an
   extended emission line region (EELR, $\sim170$\,kpc$^2$) in the
   disk of the less massive companion (ShaSS\,622). New integral-field
   spectroscopy and the multi-band data-set, spanning from
   far-ultraviolet to far-infrared and radio wavelengths, allowed us
   to obtain a detailed description of the ShaSS\,622-073 system. The
   gas kinematics shows hints of interaction, although the overall
   velocity field shows a quite regular rotation in both galaxies,
   thus suggesting that we are observing their first encounter as
   confirmed by the estimated distance of 21\,kpc between the two
   galaxy centers. The detected $\sim 2-3$\,kpc AGN outflow and the
   geometry of the EELR in ShaSS\,622 support the presence of a hollow
   bi-cone structure. The status and sources of the ionization across
   the whole system have been analysed through photoionization models
   and a Bayesian approach which prove a clear connection between the
   AGN and the EELR. The luminosity of the AGN ($2.4\times
   10^{44}$\,erg\,s$^{-1}$) is a factor 20 lower than the power needed
   to excite the gas in the EELR (4.6$\times10^{45}$\,erg\,s$^{-1}$)
   indicating a dramatic fading of the AGN in the last
   $3\times10^4$\,yr. ShaSS\,073-622 provides all the ingredients
   listed in the recipe of a {\it light echo} where a high-ionised
   region maintains memory of a preceding more energetic phase of a
   now faded AGN. This is the first case of a {\it light echo}
   observed between two galaxies.

\end{abstract}




\section{Introduction} 
\label{sec:intro}

A comprehensive and all-purpose description of the evolution of
galaxies is one of the most ambitious goals of astrophysicists
and a rewarding approach in this investigation consists in the
selection and study of samples of galaxies which are experiencing a
specific transformation through tidal interaction, ram-pressure
stripping and the like -- the so called {\it smoking guns}. Signatures
of an ongoing transformation may be detected in the galaxy colors,
star-formation indicators, morphologies and kinematics
\citep[e.g][]{DRD15,SZS16,C14,RWK99}, so their detection encompasses the
whole spectral range and galaxy regions recommending a multi-band
approach together with integral-field spectroscopy (IFS) to
effectively understand the physics of the mechanism/s at work either
internal or related to the galaxy environment.

One of the features that is more likely associated to an ongoing
transformation is the presence of extra-planar ionized gas. This can
be driven by ram-pressure stripping \citep[RPS,][]{GG72},
galaxy-galaxy interactions \citep{TT72} or nuclear activity
\citep{SFC06}. In particular, several active galaxies show regions of
ionized gas extending out of their disk \citep[e.g.][]{FS09} whose
geometry, structure, extent and luminosity provide unique information
about the physics of the active galactic nuclei (AGNs) and their
interaction with the host galaxy \citep[see][for a review]{KP15}.

In broad terms, clouds of ionized gas detected in active galaxies
either are classified as narrow-line regions (NLRs), having different
sizes and morphologies, but generally having a bi-cone structure as
predicted by the unified AGN scenario \citep[see][]{N15}, or as
extended emission line regions (EELRs) which are likely produced in
merger events, and which are dynamically decoupled from the host
galaxy.  Such regions may be either photo- or shock-ionized.

Although the physics of the AGN has been deeply investigated
\citep[see][for a review]{F12}, the complexity of this process, which
involves the whole properties of the interstellar medium (ISM, dust
content and chemical abundances) and stellar populations of the host
galaxy, may give rise to very different effects. Extended NLRs (ENLRs)
are rather common in luminous quasars \citep[e.g.][]{HHG14}, however
at lower redshifts few restricted samples of lower luminosity objects
presenting ENLRs were recently available to gain new insights into the
complex field of AGN variability and structure.

A very specific example of this complexity is the peculiar object
nicknamed Hanny's Voorwerp, a cloud of ionized gas located 20\,kpc
outside the spiral galaxy IC\,2497 at redshift $z\sim 0.05$
\citep{LSK09}.  This has been interpreted as a light echo ionized by
an AGN \citep{RGJ10} which faded dramatically within the last
$\sim10^5$\,yr. This was the first evidence for such a time scale of
AGN variability. To explain the nuclear starburst of IC\,2497
\citep{RGJ10}, the huge reservoir of H\,{\sc i} around the galaxy
\citep{JGO09}, the discrepancy between the level of ionization of the
cloud and both the luminosity of the AGN and the nature of the X-ray
emission \citep{LSK09,SSK16}, \citet{KLS12} suggested the following
sequence of events. A major merger produced the massive tail of
H\,{\sc i} surrounding both the galaxy and Hanny's Voorwerp and
triggered the AGN activity. Later, the ionizing luminosity from the
AGN activity dropped abruptly leaving Hanny's Voorwerp as the only
highly ionized region in the system.

The discovery of this system boosted a systematic search for similar
objects. \citet{KCB12} identified 19 galaxies with AGN-ionized regions
at projected radii $r_p > 10$\,kpc out of 18,166 candidates drawn from
the SDSS DR\,7. Among these, 8 show a strong deficit in
ionizing luminosity as well as no evidence of strongly obscured
AGN. Almost all of these (probably faded) AGN candidates present signs
of interaction or merger and the authors claimed that the observed
EELRs are largely photoionized tidal debris. Nevertheless, none of
these object has been caught in the act of interacting or merging with
a companion. Thus, we are observing the results and not the catalyst.

\citet{SDH13} discovered a class of 29 Seyfert-2 galaxies in the
redshift range $z=0.2-0.6$ from the SDSS DR\,8. These are
characterized by bright NLRs, possibly surrounded by ENLRs. Due to
their high-luminosity [O\,{\sc iii}] $\lambda$5007 emission, these
objects have received the nickname {\it green beans} (GBs) similarly
to the emission-line compact galaxies found by \citet{CSS09} which
have been dubbed {\it green peas}. Although having [O\,{\sc iii}]
luminosities two orders of magnitude higher with respect to the Keel
et al.'s sample and spanning a different redshift range, they share
the property of that sample in that the AGN luminosity is insufficient
to power the ENLRs, and in particular the [O\,{\sc iii}] flux. This
suggests that these objects as well may be classified as AGN
ionization echoes. \citet{SDH13} have also found hints of interaction
with neighbouring galaxies in their GBs sample.

Objects such as Hanny's Voorwerp or the GBs are rare. This can be
either due to the intrinsic low probability of such phenomena or to
their short duration. However, with respect to IC\,2497, \citet{KLS12}
pointed out that it is ``unlikely that a very rare event would be
represented so close to us'' which rather suggests that this is a
common phenomenon in AGNs. Therefore, detailed study of these few
cases is required in order to improve our understanding both of AGN
physics and its impact on galaxy evolution.

In this work, we study a galaxy pair in a very early phase of
interaction. The more massive galaxy has a Seyfert-2 nucleus which is
illuminating a vast area of the disc of the companion galaxy.  Through
a detailed analysis of the properties of this system we aim at
shedding light on the origin and nature of the EELRs. This galaxy
pair, at $z\sim0.05$, belongs to a relatively sparse region in the
Shapley Supercluster. Our analysis is based on new IFS observations
and complementary multi-band data from the Shapley Supercluster Survey
\citep[ShaSS,][]{ShaSSI}, as well as from literature.

We describe the selection and properties of the target in
Sect.~\ref{sec:target} and the IFS observations in
Sect.~\ref{sec:WiFeS}, then in Sect.~\ref{sec:SCM} the modelling of
the stellar component of the spectrum is briefly explained. Gas
kinematics and physical properties are derived and analyzed in
Sects.~\ref{sec:gaskin} and \ref{sec:gasphys},
respectively. Photoionization modelling is detailed in
Sect.~\ref{sec:models}, while the structure of the system is described
in Sec.~\ref{sec:struct}. We discuss our results in
Sect.~\ref{sec:dis} and summarize them in Sect.~\ref{sec:CONC}.

Throughout the paper we adopt a cosmology with $\Omega_M$=0.3,
$\Omega_\Lambda$= 0.7, and H$_0$=70\,km\,s$^{-1}$Mpc$^{-1}$. According
to this cosmology 1\,arcsesc corresponds to 0.96\,kpc at the target
redshift.

\begin{figure}
\includegraphics[width=80mm]{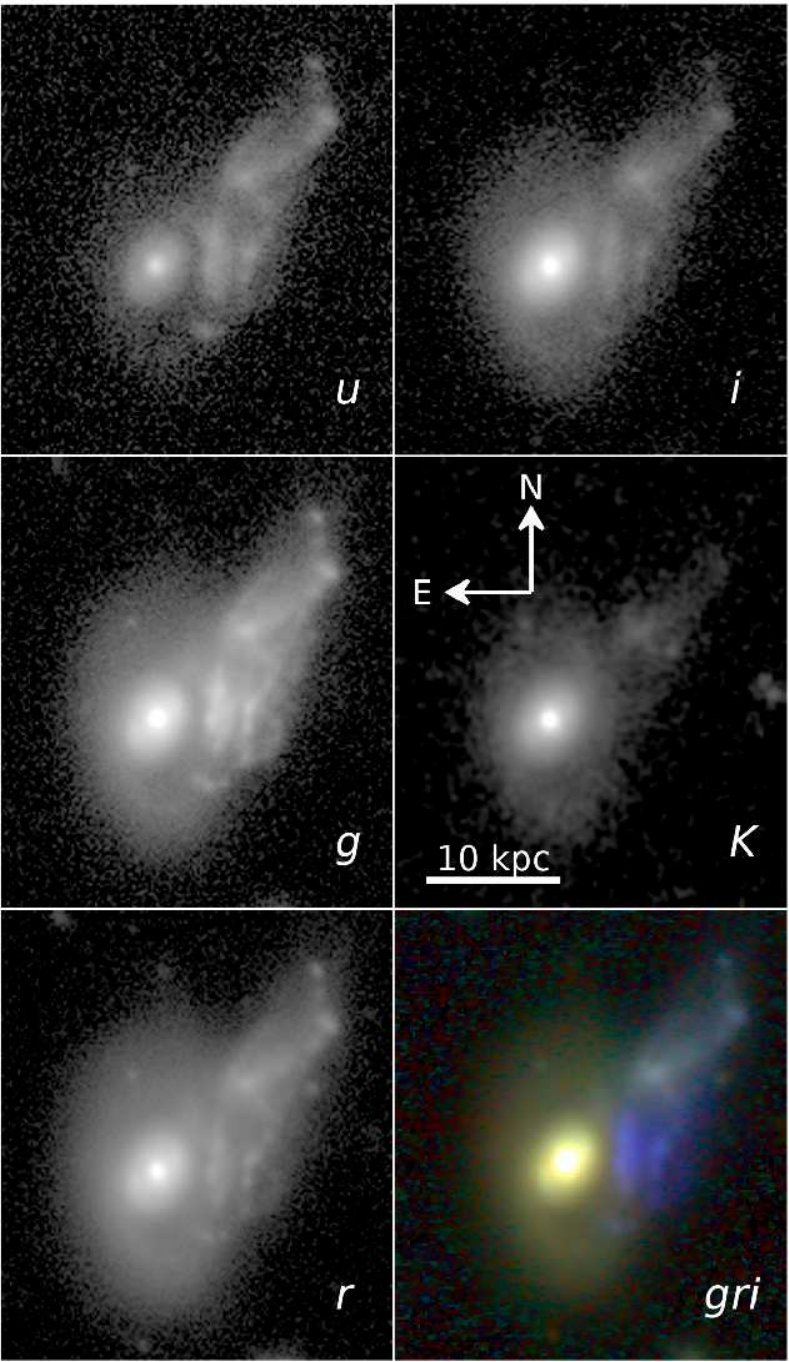}
\caption{From top to bottom and from left to right: VST $ugri$ bands
  and VISTA $K$-band images of the ShaSS\,073-622 system. ShaSS\,073 is the
  brightest galaxy near the center of the frame and ShaSS\,622 is the
  gas-rich galaxy NW to it.  Bottom right: color composite image with VST $gri$ bands assigned to BGR
  channels. The bright blue area indicates high $g$-band flux due to
  strong \OIII\ emission. North is up and East left in all panels. The
  orientation of this figure is adopted for all the other figures in
  this work unless differently specified.
  \label{fig:ugriKbands}}
\end{figure}

\section{The ShaSS\,073-622 system}
\label{sec:target}

\subsection{Target selection}
\label{sec:sel}

Our target has been identified as part of the Shapley Supercluster
Survey \citep[ShaSS,][]{ShaSSI}. This survey covers 23\,deg$^2$
centered on the Shapley Supercluster core ($z\sim 0.05$) and combines
observations in the ESO-VST $ugri$ filters\footnote{The VST images
  have been processed at the VST Data Center using the VST-Tube
  pipeline \citep{GCL12}.} with ESO-VISTA $K$-band imaging.  The ShaSS
imaging has a resolution $\lesssim 1$\,kpc\footnote{The FWHM of the
  optical images correspond to $\sim 0.8$, $\sim 0.7$, $\sim 0.6$,
  $\sim 0.9$ and $\sim 0.9$\,arcsec. for the $ugriK$ wavebands.} at
the supercluster redshift which allows us to study the structure of
the individual supercluster galaxies. 

We exploited the image quality to search for galaxies undergoing
transformation and to identify candidates for further IFS
observations. All the targets selected for the IFS were drawn from the
spectroscopic catalogue of ShaSS which is 80 per cent complete down to
$i = 17.6$ (m$^\star$+3). All these galaxies are supercluster members,
are fully resolved in the optical images and display disturbed
morphologies, such as asymmetry and tails, hints of extraplanar
emission as well as evidence of star-forming knots. About 80 galaxies
satisfy at least two of these criteria. After this {\it visual}
selection, the galaxies are targeted with a 45-minute exposure with
WiFeS which allows to ascertain which of them actually present
extraplanar emission and then become the high-priority targets in our
investigation. At present 17 supercluster galaxies have been
observed. They belong to different environments, from dense cluster
cores to the regions where cluster-cluster interactions are taking
place, and out to the much less populated areas.

The system ShaSS\,073-622\footnote{The system consists of the two
  galaxies ShaSS\,423045073 and ShaSS\,423045622 drawn from the
  Shapley Supercluster Survey \citep{ShaSSI,ShaSSII}. In the paper we
  use an abbreviated form of these identifications.} at $z=0.04865$
was chosen from the above sample because it appeared to be an
interacting pair with one member (ShaSS\,073) hosting a Seyfert-2
nucleus \citep{VV01}. This was considered a promising target to study
the early phase of the formation of EELRs.

The $ugriK$ ShaSS imaging of our object is shown in
Fig.~\ref{fig:ugriKbands}. In the $g$- and $r$-band images the main
galaxy, ShaSS\,073, appears as a barred galaxy with an external
(slightly asymmetric) ring \citep[e.g.][see
Sect.~\ref{sec:ugriK}]{B13}. Note that in all filters the central
galaxy presents a star-like nucleus showing that the galaxy hosts an
AGN. 

A color composite image of the ShaSS\,073-622 system with the VST
$gri$ bands assigned to BGR channels is also shown in
Fig.~\ref{fig:ugriKbands} (bottom right panel) which emphasizes the
most striking feature of this system -- the bright blue area West of
the ShaSS\,073 nucleus, due to strong \OIII\ emission (contributing to
the $g$ band at this redshift).

The orientation of Fig.~\ref{fig:ugriKbands} is adopted for all the
other figures in this work unless differently specified.

\begin{deluxetable}{lll}
\tablecaption{ The galaxies \label{tab:gals}}
\tablecolumns{3}
\tabletypesize{\scriptsize}
\tablewidth{60pt}
\tablehead{
\colhead{{\bf Property}} & \colhead{{\bf ShaSS\,423045073}} & \colhead{{\bf ShaSS\,423045622}} \\
}
\startdata
{\bf Coordinates}\tablenotemark{1,2} & 13\,16\,32.58\,-31\,12\,18.5 & 13\,16\,32.02\,-31\,12\,11.5 \\
{\bf (J2000)} && \\
&& \\
{\bf Magnitudes}\tablenotemark{3,4,5} & & \\
$u\tablenotemark{(a)}$ & 17.70${\pm}$0.04 & 17.68${\pm}$0.04 \\
$g\tablenotemark{(a)}$ & 16.20${\pm}$0.02 & 16.64${\pm}$0.02 \\
$r\tablenotemark{(a)}$ & 15.47${\pm}$0.02 & 16.87${\pm}$0.02 \\
$i\tablenotemark{(a)}$ & 15.11${\pm}$0.02 & 17.03${\pm}$0.02 \\
$K\tablenotemark{(a)}$ & 12.22${\pm}$0.03 & 14.95${\pm}$0.03 \\
W1\tablenotemark{(b)} & 11.69${\pm}$0.02\tablenotemark{(c)} & \\
W2\tablenotemark{(b)} & 10.59${\pm}$0.02\tablenotemark{(c)} & \\
W3\tablenotemark{(b)} & 7.19${\pm}$0.02\tablenotemark{(c)} & \\
W4\tablenotemark{(b)} & 4.79${\pm}$0.03\tablenotemark{(c)} & \\
F$_{\mathrm{60\mu m}}$ & 0.40${\pm}$0.05\,Jy\tablenotemark{(c)} & \\ 
F$_{\mathrm{1.4GHz}}$ & 3.9${\pm}$0.6\,mJy\tablenotemark{(c)} & \\
&& \\
\enddata
\tablenotetext{a}{This work. Magnitudes are given in the AB photometric system.}
\tablenotetext{b}{Magnitudes are given in the Vega photometric system.}
\tablenotetext{c}{Centered on ShaSS\,073.}
\tablecomments{Sources: $^1$ \citet{ShaSSI}; $^2$ \citet{ShaSSII}; $^3$ \citet{WISE10}; $^4$ \citet{NHv84}, $^5$ \citet{CCG98}}
\end{deluxetable}

\subsection{Properties of the ShaSS\,073-622 system}
\label{sec:data}

\begin{figure}
\begin{centering}
\includegraphics[width=85mm]{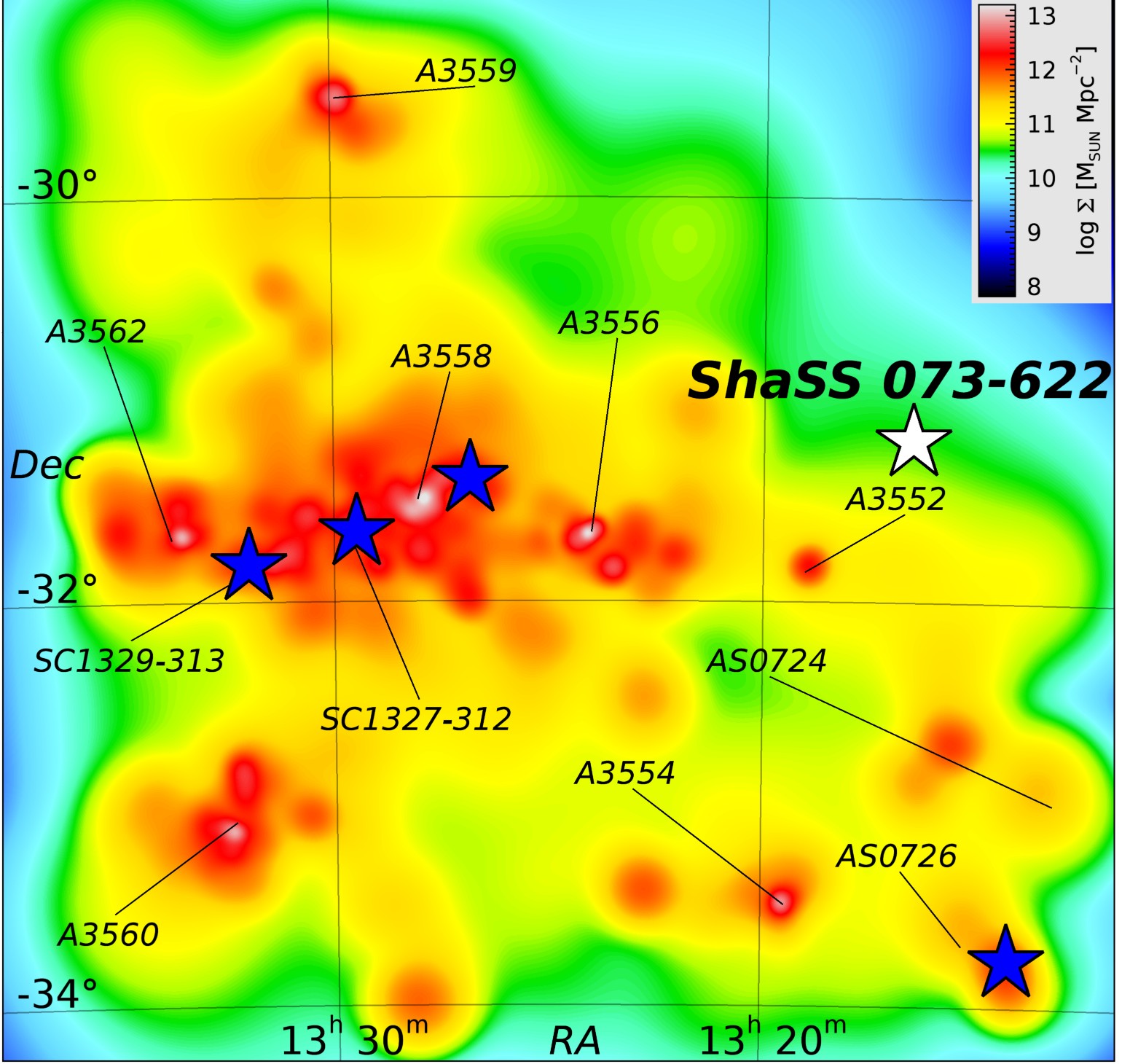}
\end{centering}
\caption{ The ShaSS stellar mass surface density map in
  M$_\odot$\,Mpc$^{-2}$ derived from the flux at 3.4\,$\mu$m
  \citep{ShaSSI}. Abell clusters and groups are labelled by black
  straight lines pointing on the X-ray center for all systems except
  AS\,0726. The upper right-hand corner is not covered by ShaSS. Blue
  stars identify the location of four galaxies in the clusters with
  ascertained ongoing RPS. The white star indicates the location of
  ShaSS\,073-622 system in a low-density environment.
\label{fig:environment}}
\end{figure}

In Fig.~\ref{fig:environment}, we show the ShaSS stellar mass surface
density map derived from the flux at 3.4$\mu$m from the WISE data
\citep[see][]{ShaSSI}. We have indicated the position of our target as
a white star together with four other supercluster galaxies with
confirmed ongoing RPS \citep[see][]{ACCESSV,ShaSSIII}. While the four
galaxies affected by RPS are associated with clusters, ShaSS\,073 is
located in a low density region. The photometric properties of the
galaxies ShaSS\,073 and ShaSS\,622 are listed in
Table~\ref{tab:gals}. The optical and $K$-band magnitudes are derived
from the ShaSS imaging as explained in Sect.~\ref{sec:ugriK}. The
projected distance between the two galaxy centers is $\sim$10\,arcsec
and the difference of their radial velocity is $\sim$130\,\kss.

The source is detected in all four channels of the Wide-field Infrared
Survey Explorer \citep[WISE,][]{WISE10}. The WISE emission in all four
bands (W1--W4) is centered on the nucleus of ShaSS\,073, suggesting
that all of the emission in these bands is due to the primary galaxy
rather than the companion. The galaxy has $\mathrm{W1-W2=1.1}$,
$\mathrm{W2-W3=3.4}$, $\mathrm{W3-W4=2.4}$, while normal quiescent or
star-forming galaxies have $\mathrm{W1-W2\sim0.0}$. \citet{SAB12} use
$\mathrm{W1-W2}>0.80$ as a criterion to select AGNs. \citet{JCM11}
select AGNs as lying within the box: $2.2<\mathrm{(W2-W3)}<4.2$,
$0.1\times \mathrm{(W2-W3)}+0.38<\mathrm{(W1-W2)}<1.7$ where is also
located our system.

\begin{figure}
\begin{centering}
\includegraphics[width=84mm]{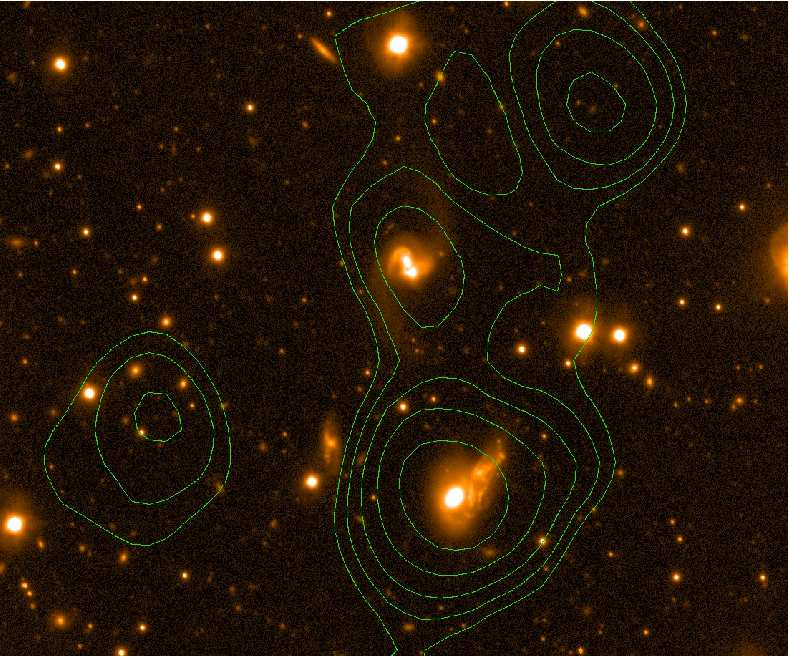}
\end{centering}
\caption{NVSS radio contours overlaid on the VST $r$-band image. The
  ShaSS\,073-622 system is in the lower part of the image, while the
  galaxy near the center is at $z\sim$0.07 and is not interacting with
  it.}
\label{fig:NVSS}
\end{figure}

The system has been detected by IRAS which measures a flux of
F$_{60\mu m}=0.40\pm0.05$\,Jy. It corresponds to source IRAS
F13138-3056 in the IRAS Faint Source Catalog \citep{MKC90}, which is
centered $\sim28$\,arcsec E of ShaSS\,073, but within the 1$\sigma$
uncertainty in position ($\sim$20\,arcsec in RA and $\sim7$\,arcsec in
Dec). GALEX NUV and FUV channels also reveal the system, tracing the
gas of ShaSS\,622.

Our source has been also detected by the NRAO VLA Sky Survey (NVSS)
which covers at 1.4\,GHz the entire sky north of $-40$\,deg
declination. In Fig.~\ref{fig:NVSS} the NVSS radio contours are
overlaid on the VST $r$-band image. These show that the radio emission
is centered on the main galaxy. The measured flux,
F$_{\mathrm{1.4GHz}}=3.9\pm0.6$\,mJy, corresponds to a luminosity of
L$_{\mathrm{1.4GHz}}=(2.20\pm0.34)\times 10^{22}$\,W\,Hz$^{-1}$. This
is insufficient for it to be classed as a radio-loud AGN
\citep{YRC01}. 

Taking into account both the IR and radio data, we note that the
ratios F$_{\mathrm{22\mu m}}$/F$_{\mathrm{1.4GHz}}$=25.9 and
F$_{\mathrm{60\mu m}}$/F$_{\mathrm{22\mu m}}$=3.986 are indicative of
an AGN \citep{DH02,ACCESSIII,DHM14}. The galaxy however, lies on the
F$_{\mathrm{70\mu m}}$/F$_{\mathrm{1.4GHz}}$ (see below) FIR-radio
relation for normal star-forming galaxies \citep{ACCESSIII}, rather
than radio-loud galaxies. This suggests that the radio emission is
mostly due to star formation. Assuming the relation
SFR=L$_{\mathrm{1.4GHz}}/(3.44\times 10^{21}
\mathrm{W}\,\mathrm{Hz^{-1}})$, then the measured flux would imply a
star formation rate SFR$=6.4\pm 1.0$\,M$_\odot$yr$^{-1}$.

We fit the far-infrared emission by the log(L$_{\mathrm{TIR}}$)=10.75
model of \citet{RAW09}. Constraining this model with the
F$_{\mathrm{60\mu m}}$ value, we predict a 70$\mu$m flux of
F$_{\mathrm{70\mu m}}$=0.491\,Jy for the whole system, and 2.4\,mJy at
1.4GHz. This model would also predict a 24$\mu$m flux of just
33.4\,mJy, which is only 28\% of the 117.5\,mJy flux level actually
observed. This shortfall suggests that much of the 24$\mu$m (or W4)
emission is due to the AGN component rather than star formation, even
if all the 60$\mu$m emission is assumed to come from star formation.

\subsubsection{Estimates of ugriK magnitudes and stellar masses}
\label{sec:ugriK}

Estimating the optical and $K$-band magnitudes of the two galaxies is
not straightforward due to their spatial overlap which affects a large
fraction ($\sim$40\%) of their disks. To separate the flux of
ShaSS\,073 from the companion, we take advantage of the sharp division
between the two gaseous discs (see Sects.~\ref{sec:gaskin},
\ref{sec:gasphys}, \ref{sec:models} and
Figs.~\ref{fig:fluxratioNII_Ha}, \ref{fig:KBPT}).

To make the measurements of fluxes easier, we isolated a frame of
40\,arcsec side around the two galaxies, in such a way that there are
very few and faint external sources, which are easily masked out
(although contributing very little to the total counts). A line
through the center of ShaSS\,073 and oriented at P.A.$\sim13^\circ$
was adopted to separate the western half of ShaSS\,073 which overlaps
with the companion from the virtually unperturbed eastern half. We
then derived the magnitude of ShaSS\,073 integrating the counts after
masking out the western side of the frame with respect to the above
line. The total flux from the galaxy was set to twice that
flux. Other, more `refined' masking shapes lead to changes at the
level of 0.01\,mag which are certainly small considering our ignorance
of the `hidden' part of the galaxy. For ShaSS\,622, we integrated the
counts of the whole frame and subtracted the total counts estimated
for ShaSS\,073. With this method we estimated the magnitudes in the
$u$, $g$, $r$, $i$ and $K$ bands given in Table~\ref{tab:gals}.

The resulting colors for ShaSS\,073 are $u-g=1.50$, $g-r=0.73$ and
$r-i=0.36$, which are consistent with a S0-Sa galaxy
\citep[see][]{FSI95}. Taking into account the presence of the bar and
the external ring, we classify ShaSS\,073 as (R)SB0a
\citep[e.g.][]{B13}. The colors of ShaSS\,622 ($u-g=1.04$, $g-r=-0.23$
and $r-i=-0.16$) are clearly determined by the illumination from the
AGN. If we consider only the areas outside of the ionization cone (see
Sect.~\ref{sec:gasphys}) the colors become $u-g=1.35$, $g-r=0.63$ and
$r-i=0.27$, which are typical of intermediate/late spirals (Sb/Sbc)
\citep{FSI95}.

\begin{figure}
\begin{centering}
\includegraphics[width=80mm]{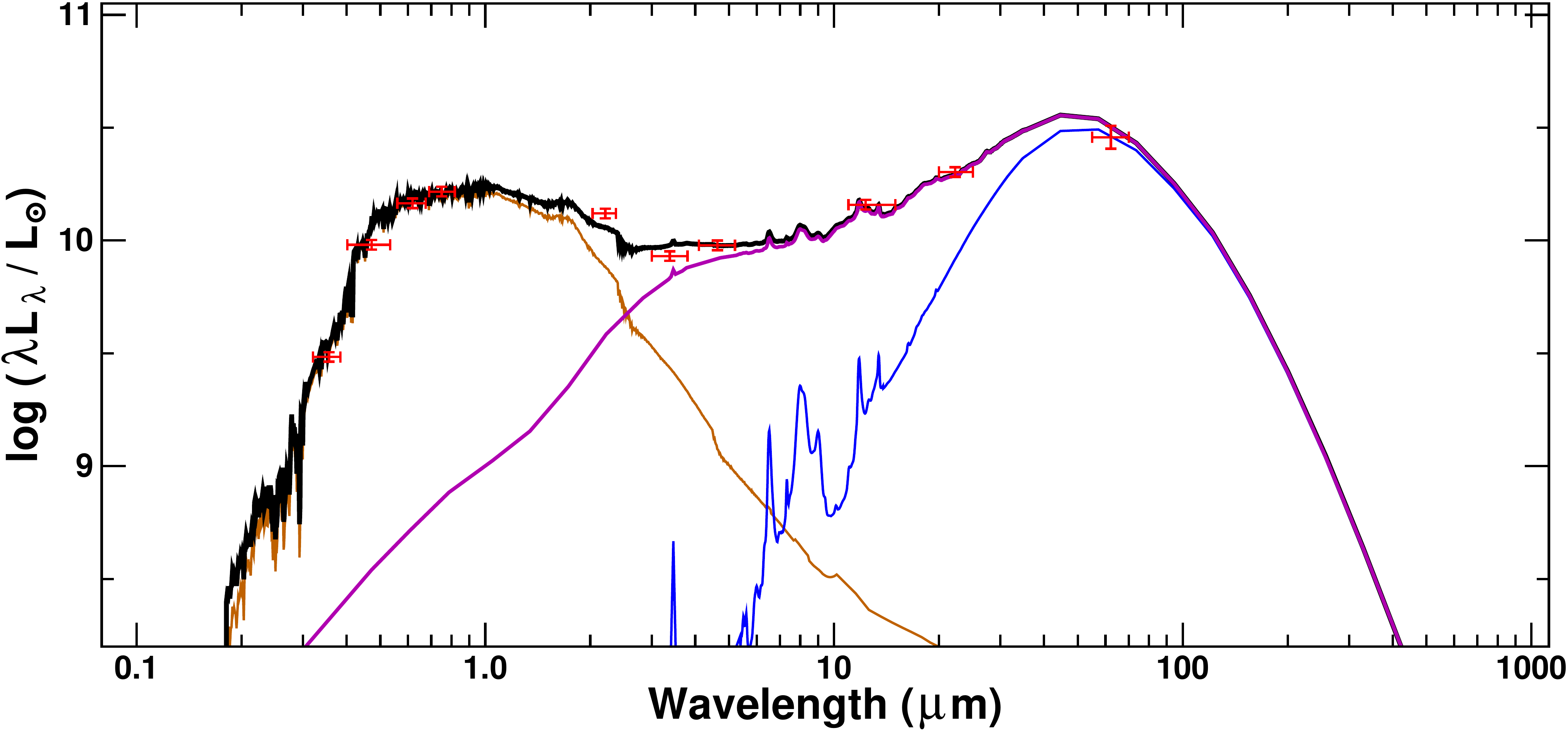}
\end{centering}
\caption{{The integrated spectral energy distribution of ShaSS\,073
    (red points with error bars) combining VST optical, VISTA
    $K$-band, WISE W1--W4 bands and IRAS 60$\mu$m flux
    measurements. The black curve shows the best-fit model SED,
    consisting of a template elliptical galaxy SED (brown curve) from
    \citet{PTM07}, combined with the two-parameter infrared SED from
    \citet{DHM14}, which includes emission from both an AGN and
    star-formation (magenta curve). The blue curve shows the
    contribution from just the star formation component in this mixed
    model.}
\label{fig:SEDfit}}
\end{figure}

From the above $g$ and $i$ magnitudes and with the adopted cosmology,
we estimate for ShaSS\,073 a stellar mass of
M*=5.7$\times$10$^{10}$\,M$_{\odot}$ adopting the calibration of GAMA
\citep{THB11}. In doing this, we neglect the AGN contribution to the
$g$ and $i$ fluxes, as suggested by the spectral energy distribution
(see Fig.~\ref{fig:SEDfit}).

The magnitudes and colors of ShaSS\,622 are affected by the
illumination from the AGN and do not reflect the properties of the
stellar populations, so that we cannot use them to estimate the
stellar mass. Instead, for the morphological types of our galaxies and
their typical colors, we can assume that their $K$-band mass-to-light
ratios are similar, according to Bell \& de Jong (\citeyear{BdJ01},
see their Fig.~3), therefore we adopted the ratios of $K$-band fluxes
as a proxy of their mass ratio. We first removed the AGN contribution
from the $K$-band flux of ShaSS\,073 as follows. From the spectral
energy distribution presented in Fig.~\ref{fig:SEDfit}, the AGN
contributes $\sim$30\% of the flux at $\sim$2\,$\mu$m. To account for
this, we multiplied by 0.7 the flux in a circle of diameter twice the
FWHM (0.9\,arcsec) of the $K$-band image, obtaining a total $K$-band
magnitude of 12.41\,mag. From the difference of the $K$-band
magnitudes we then derived a stellar mass ratio of
M*$_{073}$/M*$_{622}$$\sim$10. This classifies the potential merger
between our two galaxies as a {\em minor} or {\em intermediate} merger
\citep{HSC09,WHA12}.

\bigskip
\bigskip

\subsection{Spectral energy distribution of ShaSS\,073}
\label{sec:SED}

Fig.~\ref{fig:SEDfit} shows the spectral energy distribution of
ShaSS\,073 with measured luminosities in ten optical and infrared
bands (red points): the $ugri$ photometry from VST, the VISTA $K$-band
photometry, the WISE W1--W4 bands and the IRAS 60$\mu$m band. The
GALEX photometry is not used as the NUV and FUV emission spatially
coincides with the companion galaxy and the ionized gas cloud rather
than ShaSS\,073.

The mid-infrared continuum can be well described as a featureless
power-law over 3--60\,$\mu$m, consistent with dust heating powered by
nuclear activity \citep{SEG05,DKB12}.

The brown curve shows the best-fit template elliptical galaxy SED from
\citet{PTM07}, which reproduces well the observed optical photometry
as well as the absence of any significant ultraviolet emission from
ShaSS\,073 in the GALEX images. This leaves little scope for
unobscured emission from star formation or an AGN. We instead fit the
infrared photometry using the two-parameter semi-empirical model SEDs
of \citet{DHM14}, which include contributions from both an AGN and
normal star formation. The infrared emission from star formation is
modelled using slightly modified versions of the \citet{DH02}
templates, which represent emission from dust exposed to a wide range
of heating intensities $0.3{\le} {\mathcal U} {\le}10^{5}$ (where
${\mathcal U}=1$ corresponds to the interstellar radiation field in
the solar neighbourhood). The global dust emission of the galaxy is
then modelled as a power-law distribution of dust mass ($M_{d}$)
heated by different radiation intensities, $dM_{d}{\propto} {\mathcal
  U}^{-\alpha_{SF}}d {\mathcal U}$, giving a series of templates
parametrized by an exponent $\alpha_{SF}$ in the range 0--4. A second
parameter is introduced to account for the infrared emission from an
unobscured AGN component. For this purpose, the median mid-infrared
spectrum of \citet{SHA13} is used, which combines {\em Spitzer}
Infrared Spectrograph (IRS) observations and MIPS photometry (24, 70
and 160$\mu$m) of Palomar-Green quasars, and carefully removes any
far-infrared contribution from star formation. \citet{DHM14} produce a
suite of infrared SEDs in which the emission comes from a linear
combination of AGN and star-forming templates, in which the AGN
contributes between 0 and 100\% of the emission over the 5--20\,$\mu$m
wavelength range, spaced at 5\% intervals.

We fitted these mixed AGN+SF model SEDs to the observed infrared
emission from the $K$-band through to the IRAS 60$\mu$m band, after
subtracting the contribution from the best-fit elliptical template in
each pass-band. The best-fit overall SED combining the elliptical SED
and the mixed AGN+SF model is shown by the black curve in
Fig.~\ref{fig:SEDfit}. We were able to reproduce well the overall
power-law form of the infrared SED over 3--60\,$\mu$m as a model
(magenta curve) in which 85\% of the 5--20\,$\mu$m emission comes from
the AGN, and the remaining 15\% comes from a star-formation template
with $\alpha_{SF}{=}1.19$ (blue curve), indicating a significant
contribution from warm dust heated to 50--100\,K. The fraction from
star-formation cannot be much lower than 15\% as such models cannot
reproduce the 60$\mu$m flux, while reducing the contribution from the
AGN results in models that predict too little emission in the W2 band
and too much at 60$\mu$m.

The WISE 4.6$\mu$m (W2) band provides the key constraint on the
bolometric AGN luminosity. The photospheric emission from evolved
stars in the template elliptical SED (brown curve) drops rapidly in
the near-infrared above 2\,$\mu$m, and while the PAH complexes at
6--8\,$\mu$m and 12\,$\mu$m from H\,{\sc ii} regions can dominate the
mid-infrared emission in normal star-forming galaxies, the emission
linked to obscured star-formation drops off rapidly below 5\,$\mu$m
(blue curve) and is negligible in the W2 band. It is thus not possible
to fit the level of emission observed in the 4.6$\mu$m W2 band through
any combination of normal galaxy template SED and dust obscured
star-formation. Instead, the flat power-law emission from a torus of
hot dust, which absorbs the UV--optical radiation coming from the
accretion disk and re-emits in the infrared, is required to produce
the observed SED over 2--30\,$\mu$m, dominating the emission in all
four WISE bands. For the W2 band, 88\% of the modelled emission (black
curve) is predicted to come from the AGN component (magenta curve) and
just 12\% from the photospheric emission from evolved stars (brown
curve).

The total infrared luminosity integrated over 3--1100\,$\mu$m from the
best-fit AGN+SF model is $8.0{\times}10^{10\,}L_{\odot}$. The
far-infrared emission from the AGN component drops rapidly beyond
30\,$\mu$m, and so essentially all the IRAS 60$\mu$m flux is predicted
to come from obscured star-formation, thus providing a constraint on
the obscured SFR across the galaxy.

Based on the best-fit AGN SED model from \cite{DHM14}, the
contributions from the AGN at the 5\,$\mu$m luminosity ($\lambda \times
\mathrm{L_\lambda}$) accounts for the 90\% of the total galaxy
emission and is estimated as $3.09\times
10^{43}$\,erg\,s$^{-1}$. \citet{LRS15} provide a relation between the
5\,$\mu$m luminosity and the bolometric luminosity of AGNs, based on a
bolometric correction of a factor 8.0 \citep[assuming Type I AGN
  from][]{RLS06}. This gives us an estimate of the AGN bolometric
luminosity as being $2.47\times 10^{44}$\,erg\,s$^{-1}$.

The 6\,$\mu$m luminosity of the AGN component has been shown to
correlate linearly with the X-ray luminosity of AGNs
\citep[see][]{MCA15}. This lends support to the use of the 6\,$\mu$m
luminosity as a proxy for AGN bolometric luminosity. At 6\,$\mu$m 93\%
of the total luminosity of ShaSS\,073 should be coming from the AGN
component corresponding to a luminosity of $3.72\times
10^{43}$\,erg\,s$^{-1}$. The bolometric correction at 6\,$\mu$m should
be a factor 6.5 \citep{RLS06}, giving us an estimate of the bolometric
luminosity of $2.42\times 10^{44}$\,erg\,s$^{-1}$, in remarkable
agreement with the previous value.

\section{Integral-field spectroscopy: observations and data processing}
\label{sec:WiFeS}

The spectroscopic data on ShaSS\,073-622 system were obtained during
the night of 23 May 2015 using the Wide-Field Spectrograph
\citep[WiFeS,][]{Dopita07,Dopita10}. This instrument is mounted at the
Nasmyth focus of the Australian National University 2.3m telescope
located at the Siding Spring Observatory, Australia. WiFeS is an
image-slicing integral-field spectrograph that records optical spectra
over a contiguous 25$^{\prime\prime}\times$\,38$^{\prime\prime}$
field-of-view. The spectra were acquired in `binned mode', providing
25$\times$38 spaxels each of
1$^{\prime\prime}\times$\,1$^{\prime\prime}$ size. WiFeS is a
double-beam spectrograph providing independent channels for each of
the blue and the red wavelength ranges. We used the B3000 and R3000
gratings, allowing simultaneous observations of the spectral range
from $\sim $3300\,\AA\ to $\sim $9300\,\AA\ with an average resolution
of R=2900. For further details on the WiFeS instrument see
\citet{Dopita07,Dopita10}.

\begin{figure}
\begin{centering}
\includegraphics[width=80mm]{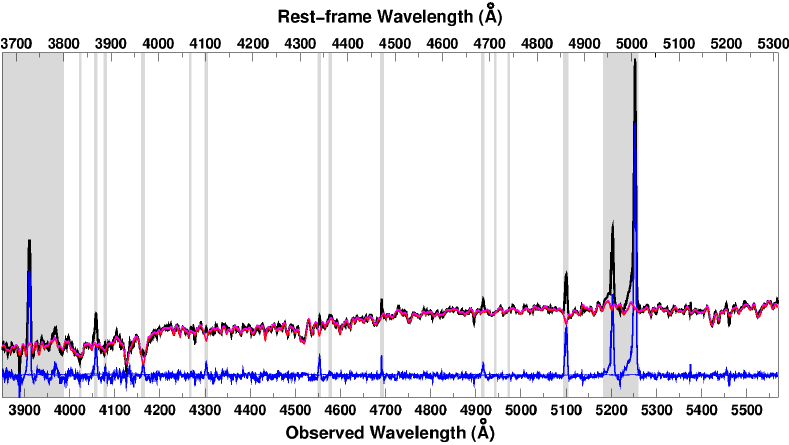}
\end{centering}
\caption{The stellar component fit to data from the blue arm of WiFeS. 
  The thick black curve shows the input spectrum, coming from a 6$\times$7
  spaxels region around the photometric center of the galaxy. The
  magenta curve shows the resultant best-fit stellar spectrum (see
  text). The shaded regions indicate the wavelength ranges excluded
  from the fitting process, including the masks for emission
  lines. The residual emission component (thick blue curve) reveals
  clear emission at [O\,{\sc ii}]\,$\lambda 3729$, H$\zeta$,
  H$\epsilon$, H$\delta$, H$\gamma$, H$\beta$ and [O\,{\sc
      iii}]\,$\lambda \lambda\,4959-5007$ in the blue arm.
\label{fig:bluespectrum}}
\end{figure}

Two exposures of 45\,min each were obtained on-target (T). We also
acquired the spectrum of a nearby sky reference region (S) with three
22.5\,min exposures in order to permit accurate sky subtraction. We
adopted the S-T-S-T-S strategy for the observation sequence. Absolute
photometric calibration of the data cubes was made using the STIS
spectrophotometric standard stars HD\,074000 and
HD\,111980 \footnote{Available at : \newline {\url
    www.mso.anu.edu.au/~bessell/FTP/Bohlin2013/GO12813.html}}. In
addition the B-type telluric standard HIP\,62448 was observed twice
during the night. Separate corrections for OH and H$_2$O telluric
absorption features were made. During the observations of the target,
the seeing averaged 0.8-1.3~arcsec, which is well matched to the
1.0~arcsec pixels of the spectrograph. Arc and bias frames were also
taken between each science exposure. Internal continuum lamp flat
fields and twilight sky flats were taken to provide sensitivity
corrections for both the spectral and spatial directions.

The data were reduced using the {\tt PyWiFeS} pipeline
\citep{Childress14}. In brief, this produces a data cube which has
been wavelength calibrated, sensitivity corrected (including telluric
corrections), photometrically calibrated, and from which the cosmic
ray events have been removed. Because the {\tt PyWiFeS} pipeline uses
a full optical model of the spectrograph to provide the wavelength
calibration, the wavelength solution is good across the whole field,
and does not rely on any interpolation of the data, since each pixel
is assigned a precise wavelength and spatial coordinate. The only
interpolation occurs when constructing the final data cube, regularly
sampled in wavelength intervals. The data achieve a SNR=5 at a flux
level of $1.0\times
10^{-17}$\,erg\,s$^{-1}$cm$^{-2}$\AA$^{-1}$arcsec$^{-2}$ at H$\alpha$.

\section{Modelling the stellar component of the spectrum}
\label{sec:SCM}

The WiFeS data allow us to derive the gas kinematics and to map
line-ratio diagnostics and dust attenuation across the galaxy
system. To achieve this we need a robust estimate of the stellar
component of the spectrum which must be identified and subtracted to
leave the pure emission-line spectrum. The details of the stellar
component modelling and emission-line measurements are fully explained
elsewhere \citep[see][]{ACCESSV,ShaSSIII}.

Briefly, we fit the stellar component from the galaxy as a linear
combination of 40 simple stellar populations (SSPs) from the
\citet{VSF10} stellar population models covering the full range of
stellar ages (0.06--15\,Gyr) and three different metallicities
[M/H]=-0.41, 0.0, +0.22. The models assume a \citet{K01} initial mass
function (IMF). They are based on the Medium resolution INT Library of
Empirical Spectra of Sanchez-Blazquez et al. (2006), have a nominal
resolution of 2.3{\AA}, close to our instrumental resolution, and
cover the spectral range 3540--7410{\AA}. The spectra were smoothed
spatially (using a $3{\times}3$ spaxel region within the main galaxy
body) to achieve a SNR of ${\sim}40$/{\AA} for the stellar continuum
at 4600--4800{\AA}. For each spaxel, the spatially-smoothed spectrum
from the blue arm was fitted with the \citet{VSF10} models after
masking out the regions below 3980{\AA}, which have significantly
reduced SNR levels and flux calibration reliability, above 5550{\AA},
where a bright sky line is located, and regions affected emission
lines. There were a dozen emission lines that required masking, and
for each line we carefully examined the data cube to identify the full
range of spectral pixels that are affected by the emission/sky line
for at least one spaxel, ensuring that the mask (which is kept fixed
for all spaxels) is sufficiently generous to account for the shifts of
the emission line due to velocity gradients. We also had to pay
particular care to fully mask the broad components linked to the
O\,{\sc iii} and H$\beta$ lines for spaxels near the nucleus of the
galaxy. For each spaxel, the best-fitting linear combination of SSPs,
recession velocity and velocity dispersion to the spatially-smoothed
spectrum is determined, and then renormalized to fit the spectrum from
the individual spaxel. This process allows us to subtract the stellar
component for spaxels in the outer regions of the galaxy where there
is a clear detection of the stellar component, but the SNR is too low
to reliably fit complex stellar population models. However, the key
aim is to reliably subtract the stellar component to enable robust
measurements of the emission lines, in particular for the Balmer lines
located in deep absorption features.

In Fig.~\ref{fig:bluespectrum} we show the best-fit linear combination
of SSPs in a central region of 6$\times$7 spaxels in ShaSS\,073. The
thick black curve is the original spectrum, the magenta curve is the
best-fit to the stellar component and the thicker blue curve is the
residual emission. The young ($< 1$\,Gyr) component of the stellar
population makes up just 4\% of the flux around 4000\AA. There is a
significant component of 2.5\,Gyr old stars, while the remainder is
8-15\,Gyr old.

\section{Kinematics of the gas}
\label{sec:gaskin}

\begin{figure}
\begin{centering}
\includegraphics[width=80mm]{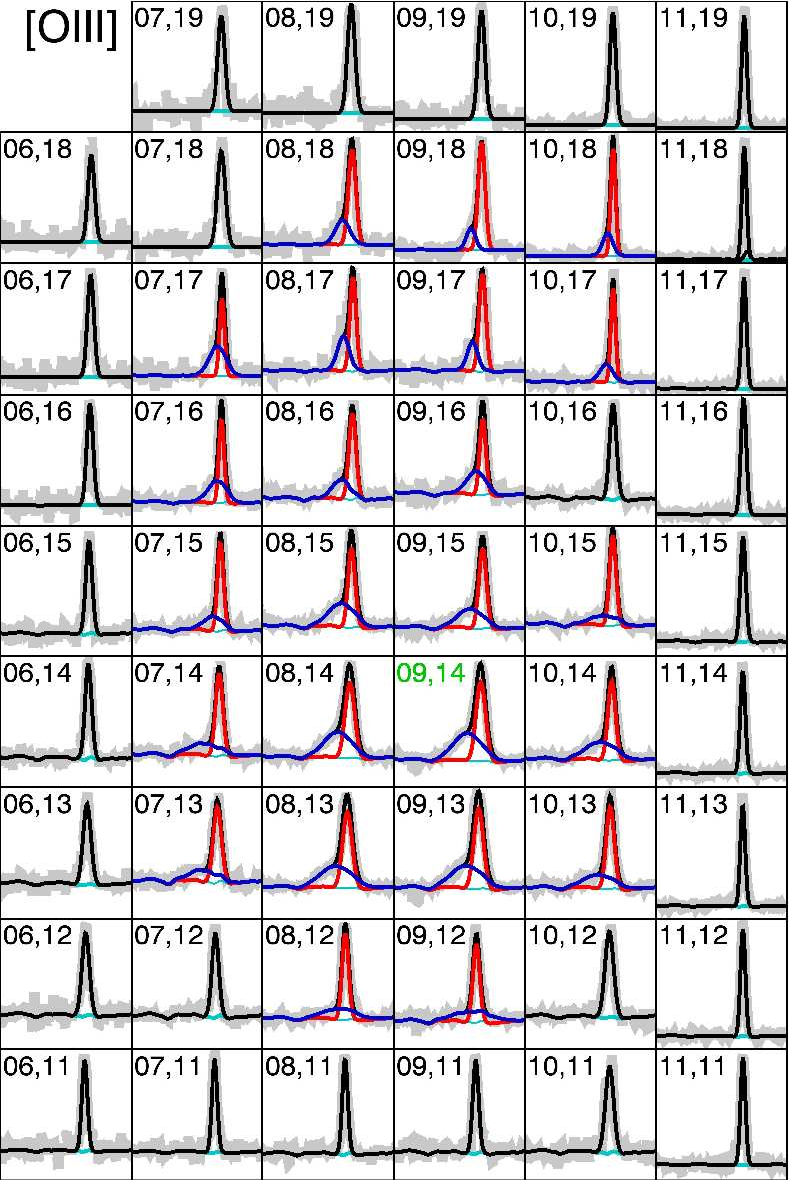}
\end{centering}
\caption{Structure of the \OIII\ $\lambda$5007 emission line in a
  region around the AGN of ShaSS\,073. Each panel shows the profile of
  \OIII\ in a 1$\times$1\,arcsec$^2$ spaxel, so that the whole figure
  covers a region of 6$\times$9\,arcsec$^2$. Pairs of numbers
  designate the coordinates in the data-cube, where (9,14), in green,
  corresponds to the center of ShaSS\,073. The curves represent: the
  observed spectrum (gray), the continuum (cyan), the one-component
  fit (black) and the two-component fit (black=red+blue). Fluxes are
  scaled to fit the height of the panels.}
\label{fig:kinmap}
\end{figure}

\begin{figure*}
\begin{centering}
\includegraphics[width=180mm]{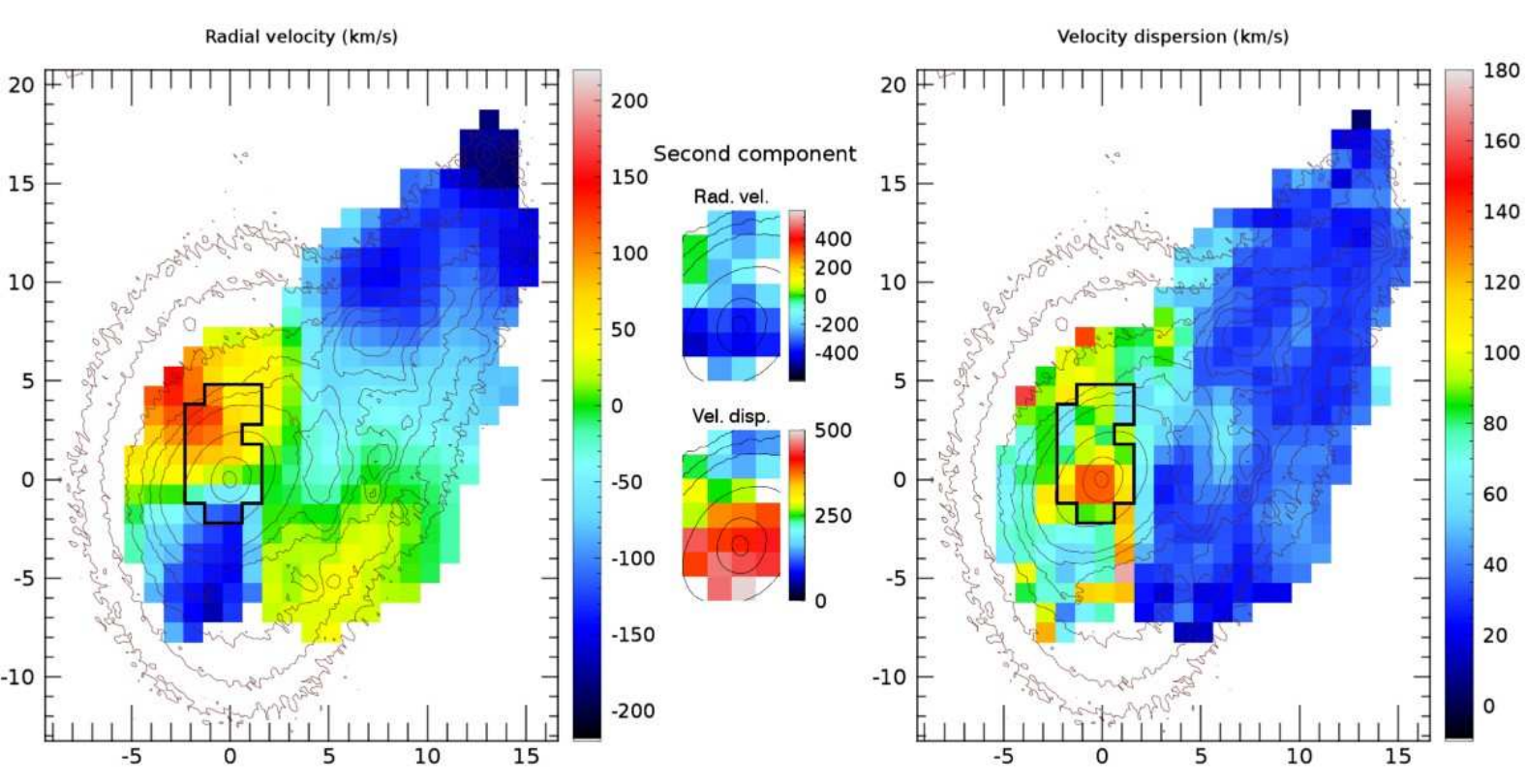}
\end{centering}
\caption{Gas velocity fields (radial velocity and velocity dispersion)
  of the ShaSS\,073-622 system in the \OIII\ line. The fields are
  shown for both kinematic components (see text and
  Fig.~\ref{fig:kinmap}). The main plots show the kinematics of the
  `main component', while the insets are for the second component
  detected in the area marked by the black polygon. Notice the
  significant differences in the scales relative to the two
  components. The contours show the $r$-band brightness
  distribution. The ticks on the plots mark the distance from the
  center of the main galaxy in arcseconds. Typical uncertainties are
  $\sim 10$\,\kss for V and $\sim 10-15$\,\kss for $\sigma$.}
\label{fig:kinetic}
\end{figure*}

The gas in the ShaSS\,073-622 system is characterized by a complex
kinematics. Fig.~\ref{fig:kinmap} shows the structure of the \OIII\
$\lambda$5007 emission line in a region of 6$\times$9 spaxels around
the center of ShaSS\,073\footnote{We study the kinematics in the
  \OIII\ $\lambda$5007 line because of its brightness. We carefully
  verified that the kinematics in the other lines (e.g. H$\alpha$) is
  fully consistent with the \OIII\ line.}. Each spaxel is
1$\times$1\,arcsec$^2$ wide (0.96$\times$0.96\,kpc$^2$ in
projection). The spatial coordinates in the data-cube are indicated,
with those of the spaxel corresponding to the center of the main
galaxy (i.e. (9,14)) written in green. The spectra are plotted in
gray, while the black-only or black-red-blue curves are the fits with
one or two spectral components respectively. Emission lines were
fitted using the package LZIFU \citep{HMG16}.

In a number (N=24) of spaxels the \OIII\ line presents two
kinematically distinct components: one component (plotted in red) has
width and redshift values forming a continuous distribution with the
rest of the galaxy; the other component (blue) is blue-shifted and in
general much wider than the former. We will refer them as main and
second components respectively. The statistical significance of
decomposition in spectral fitting has been extensively discussed in
the literature \citep[e.g.][]{HMG16}. In a first instance, we applied
the likelihood ratio test to our data \citep[Eq.~1 in][]{HMG16},
obtaining however mixed results, in some cases in obvious
contradiction with reality. We thus decided to select the spectra with
double components based on visual inspection.

The gas velocity fields are shown in Fig.~\ref{fig:kinetic}. Within
the dominant galaxy the kinematic fields have fairly smooth trends,
especially in radial velocity (V$_r$). The second component, shown in
the insets, is systematically blue-shifted and has high velocity
dispersion.

\begin{figure}
\begin{centering}
\includegraphics[width=80mm]{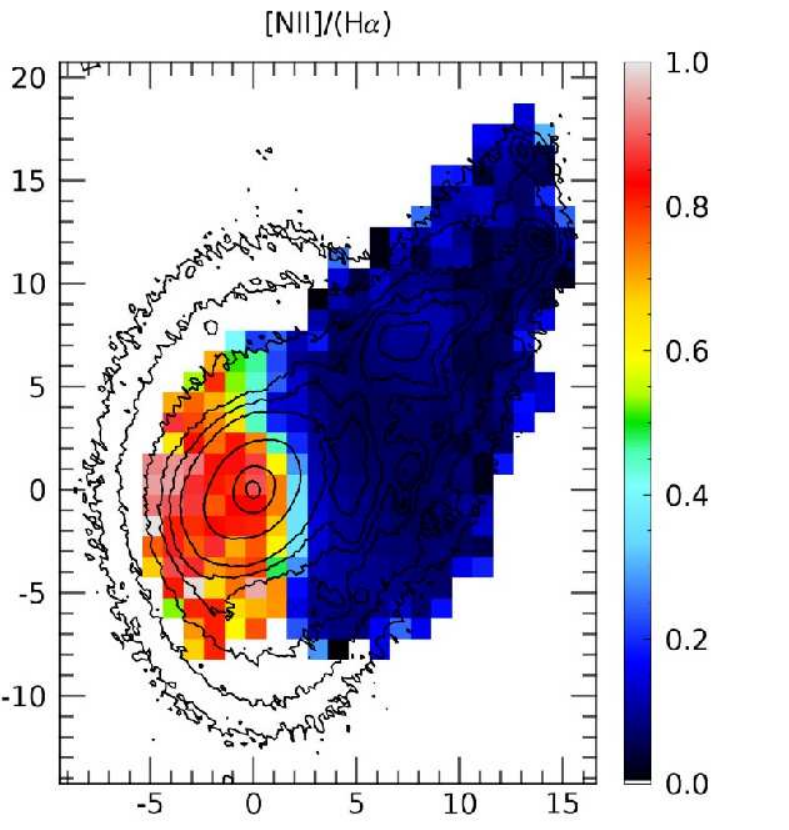}
\end{centering}
\caption{The \NII/H$\alpha$ line ratio which sharply separates the two galaxies.}
\label{fig:fluxratioNII_Ha}
\end{figure}

\subsection{Kinematics of the main gas component}
\label{sec:kmgc}

The radial velocity field of the main component gas is fairly
consistent with rotation of the gas in the disks of both galaxies. In
the main galaxy, the signature of an orderly rotational motion is
clearly visible in the eastern side of its disk (where there is no
superposition of the two galaxies). In ShaSS\,622 the rotation in the
opposite sense to the main galaxy is visible along its whole
extent. The survival of such an ordered rotation is clear evidence
that the interaction between the two galaxies is at its starting
phase. The kinematics of ShaSS\,622 will be further considered in
Sect.~\ref{sec:struct}.

To better understand the situation in the region of superposition of
the two galaxies, we show in Fig.~\ref{fig:fluxratioNII_Ha} the
distribution of the \NII/H$\alpha$ line ratio, which is a proxy of gas
metallicity. We note that in ShaSS\,622 \NII/H$\alpha$ is always lower
than $\sim$0.2, while in the main galaxy it is generally higher than
$\sim$0.6, with a very narrow ($\sim$1-2\,arcsec-wide) `transition'
strip, which is likely the superposition of the two sets of
emission. Thus, this flux ratio effectively separates the gas of the
two galaxies. A `transition strip' is clearly seen also in the radial
velocity field, with values of V$_r$ intermediate between those of the
two disks.

The separation between the two gas disk systems is also evident from
the distribution of the velocity dispersions, which are much lower
($\sim 40$\,\kss on average) in ShaSS\,622 than in the main galaxy
($\sim 60-170$\,\kss ). The mixing strip is characterised by a local
increase of $\sigma$ (the absolute maximum of $\sigma$ of 170\,\kss
actually is located in the southern strip), likely due to the
superposition of the motions of the gas in the two disks. A maximum
velocity dispersion of 135\,\kss\ is located in the center of the main
galaxy. The nucleus of ShaSS\,622 is blue-shifted with respect to that
of ShaSS\,073 by $\sim$130\,\kss, which should represent the component
along the line of sight of the relative systemic motion of the two
galaxies.

\subsection{Kinematics of the second gas component}
 \label{sec:kin2comp}

 The second gas component is generally blue-shifted with respect to
 the disk of ShaSS\,073 and has also a higher velocity dispersion. The
 value of $\sigma$ clearly divides the second component in two
 distinct areas: the circum-nuclear ($r\lesssim$2-3\,kpc) region with
 $\sigma>$360\,\kss\ reaching 490\,\kss (red-white colors in the lower
 inset of Fig.~\ref{fig:kinetic}) and radial velocity down to
 $-440$\,\kss\, and a northern area with more moderate values. We
 anticipate that this separation based on kinematics corresponds to
 distinct physical properties of the gas, as will be shown in the
 following Sections, and will postpone the discussion of the
 lower-$\sigma$ component.

\begin{figure}
\begin{centering}
\includegraphics[width=80mm]{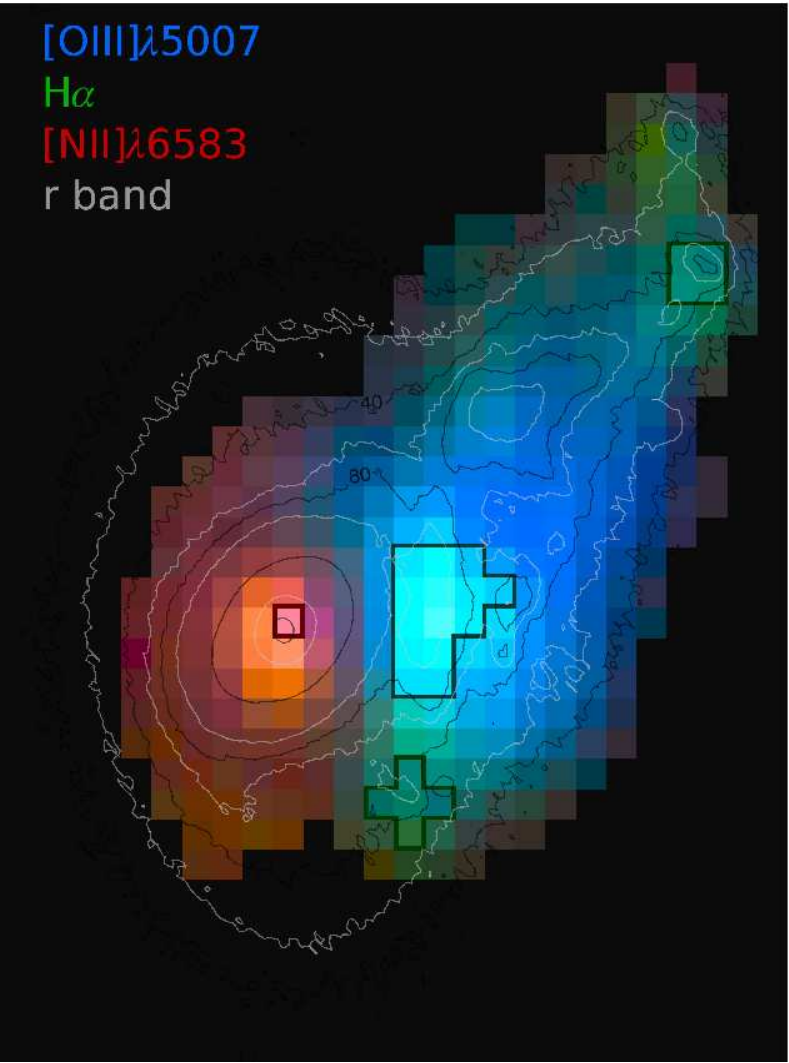}
\end{centering}
\caption{A composite RGB image derived from the fluxes in the emission
  lines \NII\ $\lambda6583$ (R), H$\alpha$ (G), and
  \OIII\ $\lambda5007$ (B), with isophotes from the VST $r$-band image
  superimposed. The black polygons mark the groups of spaxels whose
  spectra were combined to produce the representative spectra shown in
  Figs.~\ref{fig:AGN_spectra} and \ref{fig:HII}. From top to bottom
  and on the right: {\it HII Region \# 1, High-Excitation Region, HII
    Region \# 2 }. On the center left: {\it AGN}.}
\label{fig:WiFeS}
\end{figure}

The kinematics of the circum-nuclear gas, i.e. large negative radial
velocities accompanied with large velocity dispersion within a scale
of a few kiloparsecs, is typical of AGN outflows
\citep[e.g.][]{FCK13,KWB16}. Outflows are quite common in type 2 AGNs
\citep[e.g.][]{HAM14,MCP15,WBS16}, and generally have the structure of
a hollow bi-cone in which the gas flows in the external corona, while
the AGN radiation escapes from the inner cone
\citep[e.g.][]{CKH00,FCK13}. In general, only the blue-shifted half of
the outflow is observed (as in our case) because the other part is
obscured by the dust in the galactic disk. The limited spatial
resolution of our IFS prevents a detailed study of the structure of
the outflow (e.g. using the models of Fisher et al. \citeyear{FCK13}
or Bae \& Woo \citeyear{BW16}). We will however be able to gain insight
into the geometry of the ionization cone using the properties of the
disk region of ShaSS\,622 affected by it (see Sect.~\ref{sec:struct}).

\section{Physical properties of the gas}
\label{sec:gasphys}

Fig.~\ref{fig:WiFeS} shows a RGB composite image derived from the
fluxes in \NII\ $\lambda6583$ (red), H$\alpha$ (green), and \OIII\
$\lambda5007$ (blue). The ShaSS\,073-622 complex appears to be clearly
divided in three regions in this figure. The main galaxy is seen very
strongly in the \NII\ $\lambda6583$ line, indicating a high chemical
abundance, probably along with a high reddening (see below). Not
surprising for an AGN, the nucleus is very strong in this line. The
green areas at the extremes of the disk of ShaSS\,622 are constituted
by individual \HII\ region complexes, as is apparent comparing the
positions of the isophotes with the images in
Fig.~\ref{fig:ugriKbands}.

\begin{figure}
\begin{centering}
  \includegraphics[width=80mm]{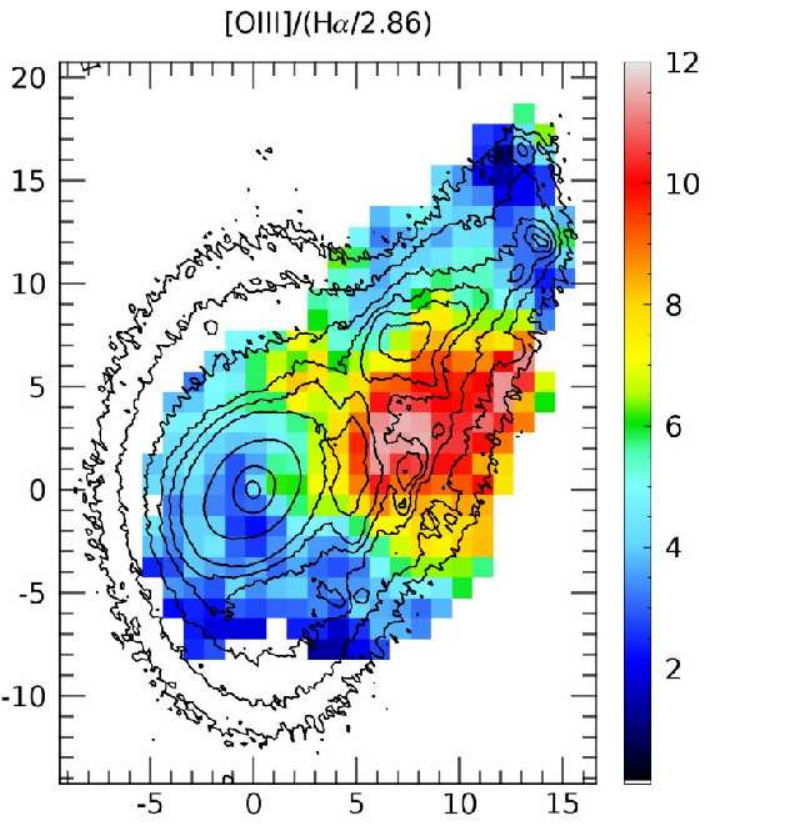}
\end{centering}
\caption{The `effective' \OIII/H$\beta$ line ratio (derived from the
  reddening uncorrected \OIII\ and H$\alpha$ line flux maps). Note the
  highly excited region in the disk of ShaSS\,622.}
\label{fig:fluxratioOIII_Hb}
\end{figure}

As already noted, the most striking emission feature of ShaSS\,622 is
the region of very strong \OIII\ emission (blue in the figure, see
also bottom right panel of Fig.~\ref{fig:ugriKbands}). The brightest
part of this region is the one closest to the nucleus of
ShaSS\,073. To get a more precise idea on its shape, we show in
Fig.~\ref{fig:fluxratioOIII_Hb} the distribution of the `effective'
\OIII/H$\beta$ line ratio, which can be taken as a proxy for the
ionization parameter for regions excited by an AGN
\citep{Davies16}. This is taken from the \OIII\ $\lambda 5007$ flux
distribution (uncorrected for local reddening, which is however very
low, cf. Fig.~\ref{fig:attenuation}) and the \Ha flux map (likewise
uncorrected for reddening) divided by the standard Balmer ratio of
2.86, which provides an approximate H$\beta$ flux from the measured
H$\alpha$. Fig.~\ref{fig:fluxratioOIII_Hb} reveals the presence of
highly ionized gas in a well-defined area of the ShaSS\,622 disk. The
border of this area appears to be rather sharp and bears some
resemblance with a parabola. Considering that the AGN outflow was
already identified in Sect.~\ref{sec:kin2comp}, we associate this
high-excitation area with the intersection of the ionization cone from
the AGN with the gas of ShaSS\,622 and we will refer to it as
  ``the high excitation region'' (HER). The aperture and orientation
of the ionization cone will be determined in Sect.~\ref{sec:struct}.

\begin{figure}
\begin{centering}
\includegraphics[width=70mm]{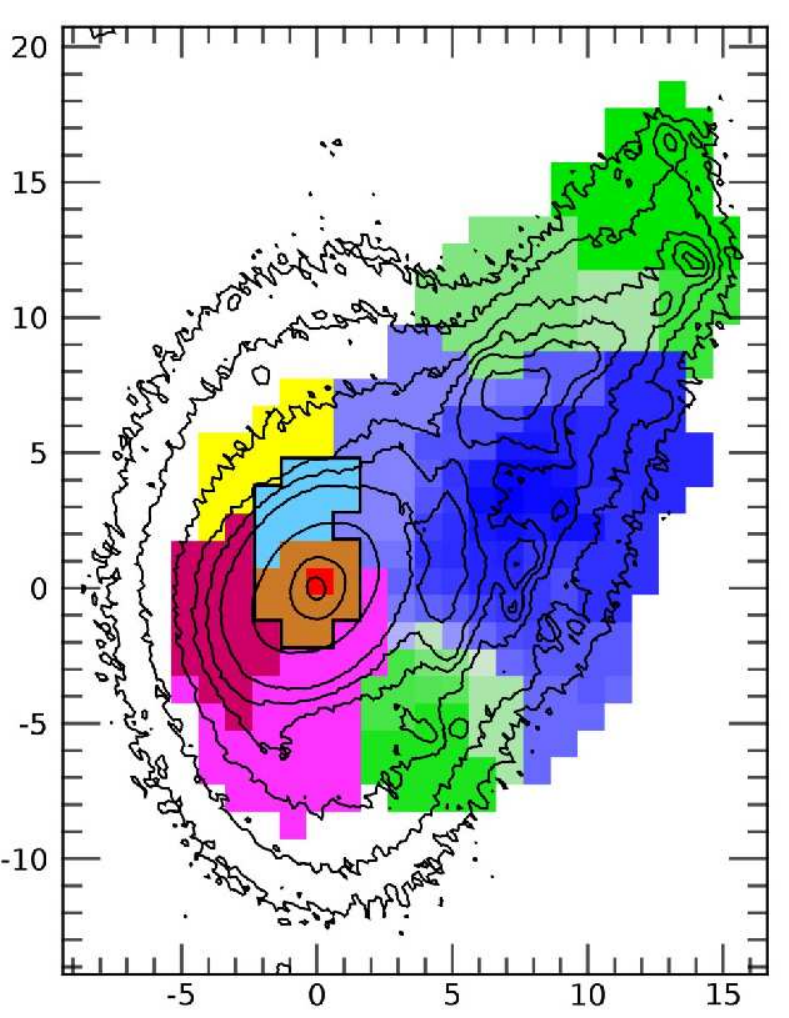}
\end{centering}
\caption{Identification of the 73 regions in which the spectra have
  been combined to derive the line ratios used in our analysis.  The
  region with a double gas component is marked with a black
  polygon. The rationale for the choice of colors is explained in the
  text and becomes clear in Fig.~\ref{fig:KBPT}.}
\label{fig:bins}
\end{figure}

A wealth of physical and chemical properties of the ionized gas can be
derived from the ratios of emission-line fluxes. In a previous work
\citep{ACCESSV} we have shown that to obtain flux ratios with
uncertainties lower than $\sim$30\%, the individual lines must have a
signal-to-noise ratio (SNR)$\gtrsim$20. To achieve these SNRs we
spatially binned the data by means of the \emph{`Weighted Voronoi
  Tessellation'} (WVT) method described by Diehl \&
Statler\footnote{http://www.phy.ohiou.edu/diehl/WVT}
(\citeyear{DS06}). This attempts to reach a fixed target SNR in all
bins \citep[see][for details]{ACCESSV}. After some tests, we set as
our target SNR=150 for the H$\alpha$ line. To obtain a spatial mapping
consistent with the distribution of the properties of the gas outlined
above, we applied the WVT to areas clearly associated with either one
or the other galaxy.

\begin{figure}
\begin{centering}
  \includegraphics[width=80mm]{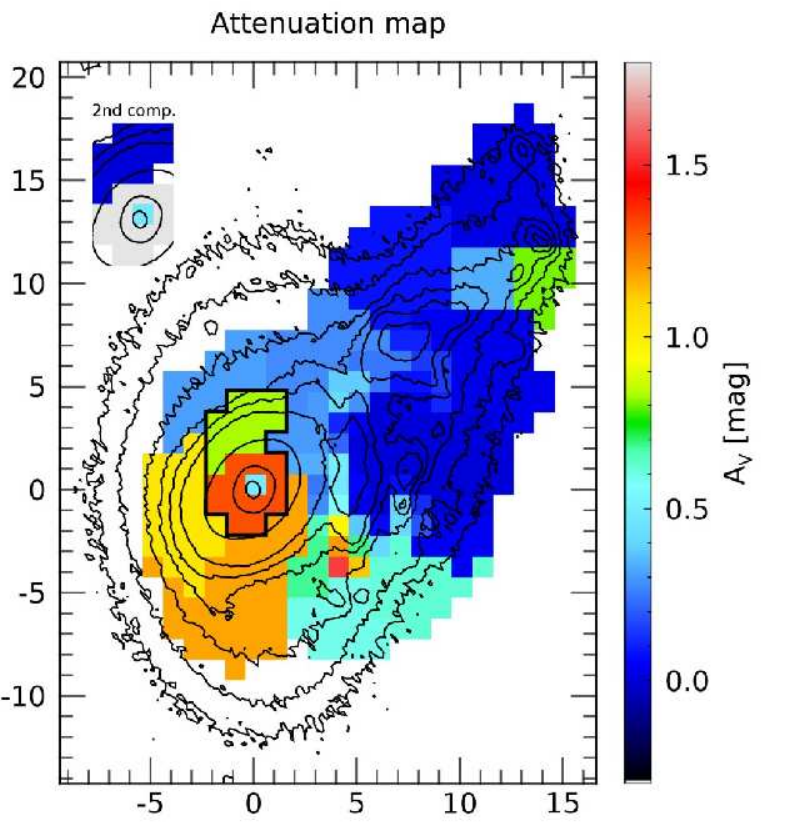}
\end{centering}
\caption{Attenuation map ($A_V$ magnitudes) derived as described in
  the text. The area where two spectral components are present is
  marked by the black polygon. The attenuation in the outflow is shown
  in the inset and reaches $A_V=3$ (light gray area).}
\label{fig:attenuation}
\end{figure}

The 73 regions identified in this way are shown in
Fig.~\ref{fig:bins}. The regions belonging to ShaSS\,622 are
color-coded according their \OIII/H$\beta$ decreasing ratios from blue
to green. ShaSS\,073 was divided in six regions: three in the disk,
one in the center (red) and two including the other spaxels with
two-component gas (orange and cyan). The orange region is where the
second component has velocity dispersion $>$360\,\kss and is more
blue-shifted (corresponding to the gas in the outflow), while in the
cyan area these features of the second component are much less
prominent (cf. Sec.~\ref{sec:kin2comp}). The colors for ShaSS\,073
were chosen for visual convenience only.

\begin{figure*}
\begin{centering}
\includegraphics[width=180mm]{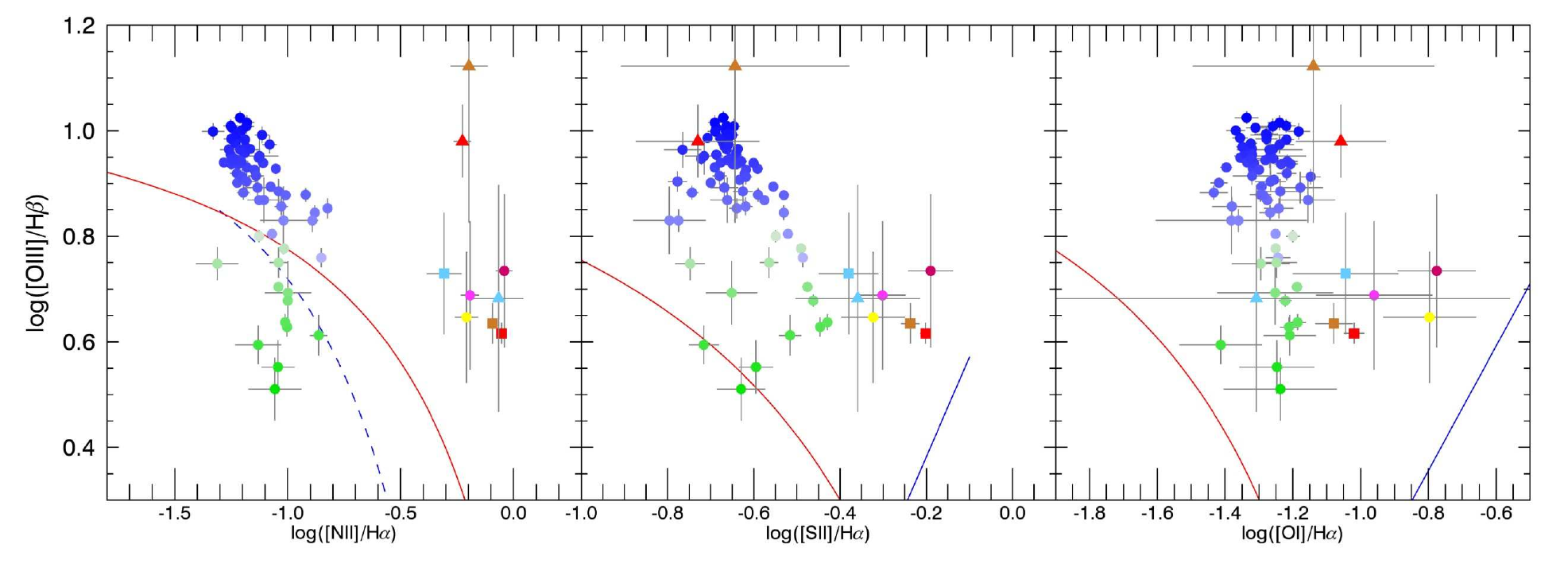}
\end{centering}
\caption{Line ratios diagnostic diagrams of the different regions of
  the ShaSS\,073-622 system. The curves are the empirical (blue
  dashed) and theoretical (red continuum) upper limits for
  \HII~regions, while the blue straight line divides Seyfert galaxies
  from LINERs (upper and lower part respectively). The colors
  correspond to Fig.~\ref{fig:bins}. For the three regions populated
  by two-component gas (with colors red, orange and cyan), square
  symbols are for the main component and triangles represent the
  second component.}
\label{fig:KBPT}
\end{figure*}

\bigskip
\bigskip

\subsection{Dust attenuation}
\label{sec:DA}
Fig.~\ref{fig:attenuation} shows the distribution of the dust
attenuation derived from the H$\alpha$/H$\beta$ line ratio, in terms
of the visual extinction $A_V$, derived in the binned regions of
Fig.~\ref{fig:bins}.  We use the theoretical attenuation curve by
\citet{FD05} with R$_V$=4.5. To derive the attenuation map we adopted
an H$\alpha$/H$\beta$ ratio of 2.86. The uncertainties on
$A_V$ are in the range 0.1-0.3\,mag over ShaSS\,622 and reach
0.5-1.1\,mag in the eastern disk of ShaSS\,073, and in the outflow,
due to lower SNR of the H$\beta$ line.

The attenuation is generally low ($<0.5$\,mag) in ShaSS\,622. It is
somewhat higher in a limited area in the NW disk ($A_V\sim0.7)$, and
in the area where the SE disk of ShaSS\,622 and the SW disk of
ShaSS\,073 overlap ($A_V\sim0.6$). The highest dust extinction takes
place in the central and SE disk of ShaSS\,073 and in the broad-line
part of the second component, where it reaches $A_V\sim 3\pm1$
magnitudes. A local minimum of $A_V\sim0.5\pm0.6$ is observed in the
nucleus of ShaSS\,073 in both the narrow and broad components.  This
fact is consistent with either dust removal driven by radiation
pressure or dust destruction from the UV radiation of the AGN.

Overall, the dust attenuation follows the general trend of the properties
of the gas, being fairly distinct between the two galaxies. The dust is
clearly more abundant in the disk of ShaSS\,073 than in the companion.

\subsection{Line ratio diagnostics}
\label{sec:lrdiag}

To study the variation in excitation and the mode of excitation of the
ionized gas in both galaxies we use the BPT emission line ratio
diagrams \citet{Baldwin81} as refined by \citet{VO87}, \citet{KHD01},
\citet{KHT03} and \citet{KGK06}.

In Fig.~\ref{fig:KBPT} we show the measured line ratios for the
different galaxy regions. The red curves mark the computed upper limit
to star-forming regions defined by \citet{KHD01}, while the dashed
blue curve in the left panel marks the empirical upper limit for
\HII~regions from \citet{KHT03}. Between the red curve and the dashed
blue curve, `composite' systems are found whose emission may be
characterised by a mixture of AGN and \HII~regions or by shock
excitation. The blue continuous line in the central and right panels
divides the line ratios typical of Seyfert galaxies (upper part of the
diagram), from those of low ionization nuclear emission line regions
(LINERs).

The regions of our system separate into distinct groups: those where
the gas tends to be mainly photo-ionized by star formation (green
dots), the HER (blue dots) which reach high levels of excitation
(log(\OIII / \Hb)$\sim1$). The third group includes all the regions of
ShaSS\,073. For the three regions populated by two-component gas
(colors red, orange and cyan in Fig.~\ref{fig:KBPT}), square symbols
are for the main component and triangles represent the second
component.

The gas in ShaSS\,073 is generally characterized by a lower excitation
than the HER (log(\OIII /\Hb)$\lesssim 0.75$), with the notable
exception of the gas in the outflow (red and orange triangles). The
second gas component in the northern disk (cyan triangle) is instead
mixed with the rest of the gas of ShaSS\,622 in the diagnostic
diagrams, showing that it belongs to the disk population and has
therefore a different nature from the outflow (see
Sect.~\ref{sec:kin2comp}). This will be further discussed in
Sect.~\ref{sec:struct}.

\section{Photoionization Modelling}
\label{sec:models}

\subsection{Analysis using Diagnostic Diagrams}
A better insight into the physics of the different regions can be
obtained by comparing the observations explicitly with the results of
photoionization modelling. In Fig.~\ref{fig:AGNgrid} we have used the
theoretical AGN grid of models from \citet{Davies16} to analyse both
the AGN and the region within the HER of ShaSS\,622. These
models were constructed using the {\tt Mappings 5.0}
code{\footnote{Available at {\tt miocene.anu.edu.au/Mappings}}
  (Sutherland et al. 2017, in prep.), and use the Local Galactic
  Concordance abundance scaling described by \citet{Nicholls17}. The
  input spectrum for the AGN is kept fixed in this grid.

Fig.~\ref{fig:AGNgrid} makes the abundance difference between the AGN
(red dots) and the HER (blue dots) in the companion galaxy very
clear. The AGN is fitted with a $\sim$1.5 times super-solar abundance
while in the HER is $\sim 0.4$ solar. The ionization parameter,
$\log U$ in the AGN is typically $-2.7$, while in the outflow much
higher values ($\log U \sim -1$) are indicated. For the HER, the
ionization parameter reaches $-1.8$, falling to $-2.5$ at the edge of
the AGN-ionized region.

\begin{figure*}
\begin{centering}
  \includegraphics[width=170mm]{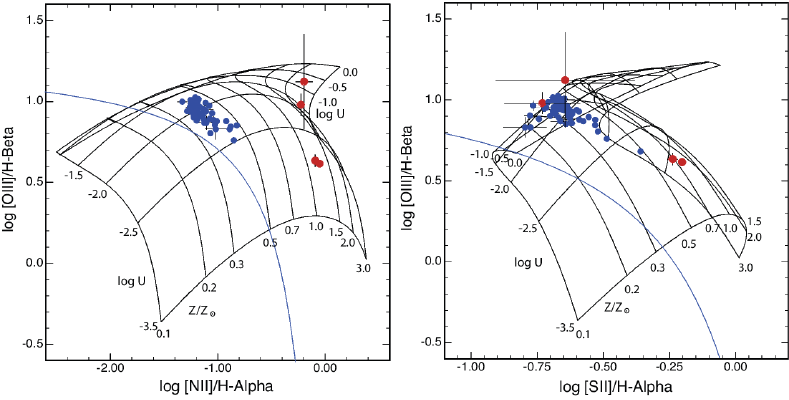}
\end{centering}
\caption{The theoretical AGN grid of models from \citet{Davies16}
  compared with the observations of the HER (blue dots) and the
  region around the AGN in ShaSS\,073 (red dots).}
\label{fig:AGNgrid}
\end{figure*}

\begin{figure*}
\begin{centering}
  \includegraphics[width=180mm]{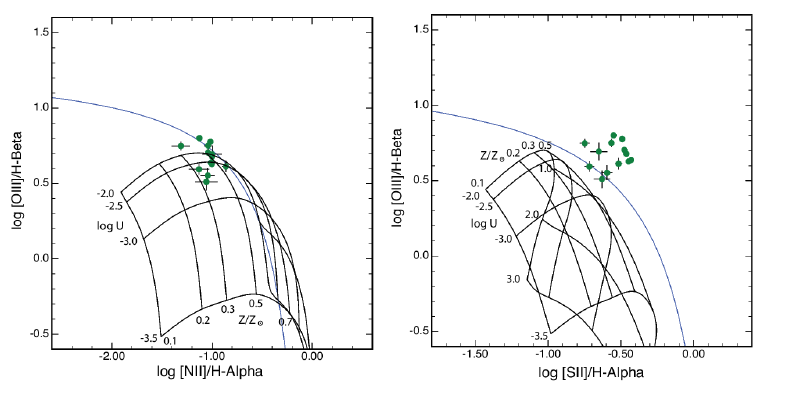}
\end{centering}
\caption{The theoretical grid of HII region models from
  \citet{Davies16} compared with the observations of the regions of
  ShaSS\,622 shown in green in Fig.~\ref{fig:bins}. The low
  metallicity inferred for the HER in
  Fig.~\ref{fig:AGNgrid} is confirmed in these star formation regions
  lying outside the ionization cone.}
\label{fig:HIIgrid}
\end{figure*}

The analysis of the `normal' star-forming regions of ShaSS\,622 (the
green bins in Fig.~\ref{fig:bins}) requires that we compare these with
\HII\ region models. This is shown in Fig.~\ref{fig:HIIgrid}. For these
we use the same approach as described by \citet{Dopita13} with input
EUV spectra derived from the {\tt Starburst 99} code \citep{SB99}. The
grid shown is for a high pressure ISM with $\log(P/k) = 6.8$
(cm$^{-3}$K).

In Fig.~\ref{fig:HIIgrid}, it is evident that the \HII\ regions in the
\NII/\Ha\ diagnostic are consistent with a metallicity of $0.3-0.7$
solar and a high ionization parameter. However, the
\SII/\Ha\ diagnostic has the observed points lying off the grid. Both
these facts, and the high ionization parameter implied by the
\NII/\Ha\ diagnostic could be explained by contamination of the bins
lying outside the main ionization cone by AGN photons penetrating the
diffuse ISM of ShaSS\,622.

The parameters of photoionized regions derived from BPT diagrams alone
are limited by the presence of other factors influencing both the
position and shape of the theoretical grids. These include the gas
pressure $\log(P/k)$ (cm$^{-3}$K), the depletion factor of iron and
other elements onto grains, and the details of the shape of the EUV
spectrum.

In order to facilitate a more detailed photoionization analysis of the
AGN in the main galaxy, the high-excitation region of the companion
galaxy, and the two brightest \HII\ regions in the companion galaxy,
we extracted spectra summing over the spaxels indicated in
Fig.~\ref{fig:WiFeS}. The resultant spectra are plotted in
Figs.~\ref{fig:AGN_spectra} and \ref{fig:HII}. \HII\ region \#1
corresponds to the northern \HII\ region in the companion galaxy, and
\HII\ region \#2 is the southern object in Fig.~\ref{fig:WiFeS}.

\begin{figure*}
\begin{centering}
\includegraphics[width=170mm]{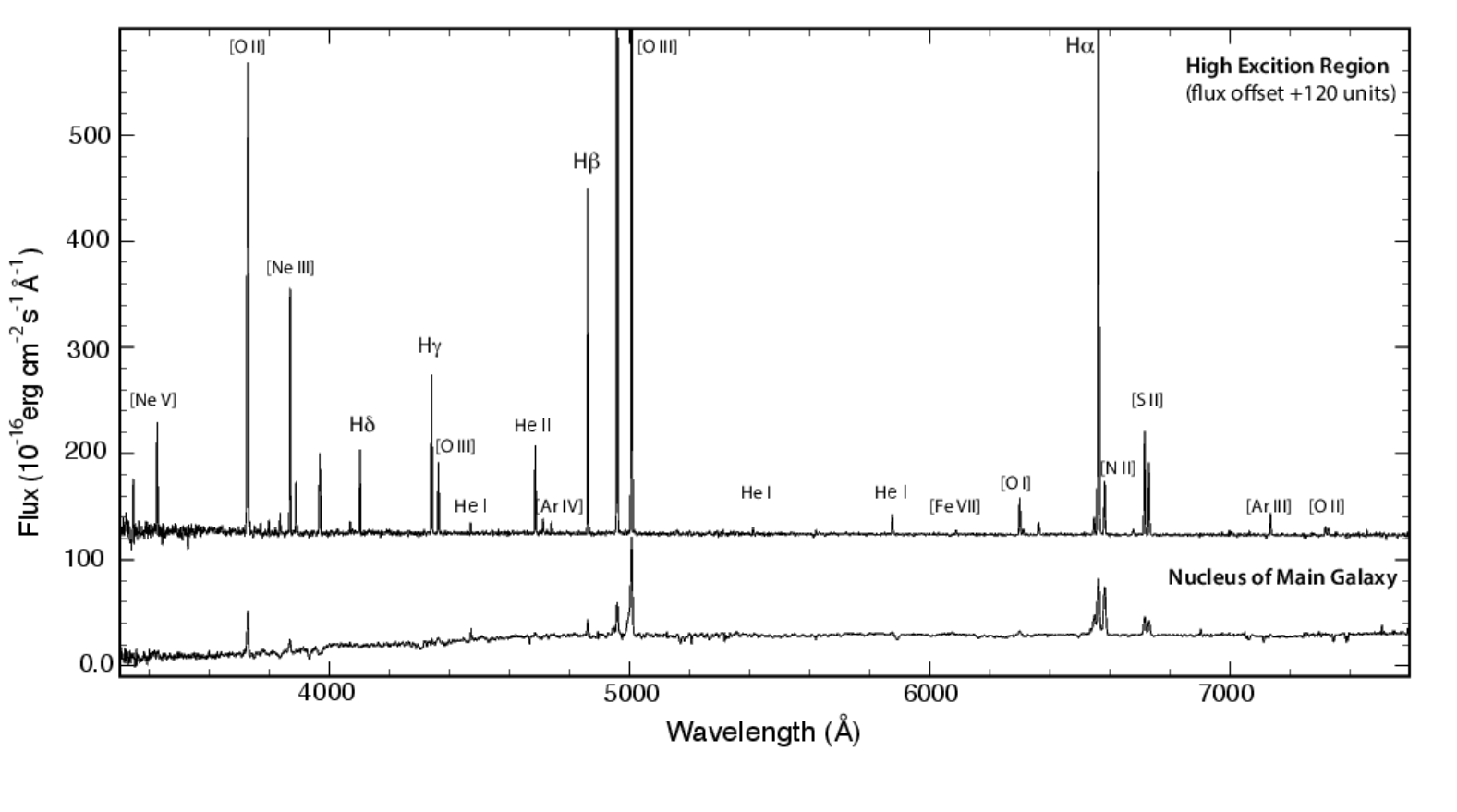}
\end{centering}
\caption{The extracted spectra of the nuclear region of the main
  galaxy and the HER. The latter spectrum has been scaled to
  show the fainter lines, so both H$\alpha$ and the \OIII\ lines are
  off scale. Note the presence of highly excited species such as [Ne
    V] and [Fe VII]. The spectrum of the nucleus of the main galaxy is
  that of a Seyfert-2 with a relatively strong continuum of old stars,
  and is the presumed source of the excitation of the high-excitation
  region of the companion galaxy.}
\label{fig:AGN_spectra}
\end{figure*}

\begin{figure*}
\begin{centering}
\includegraphics[width=170mm]{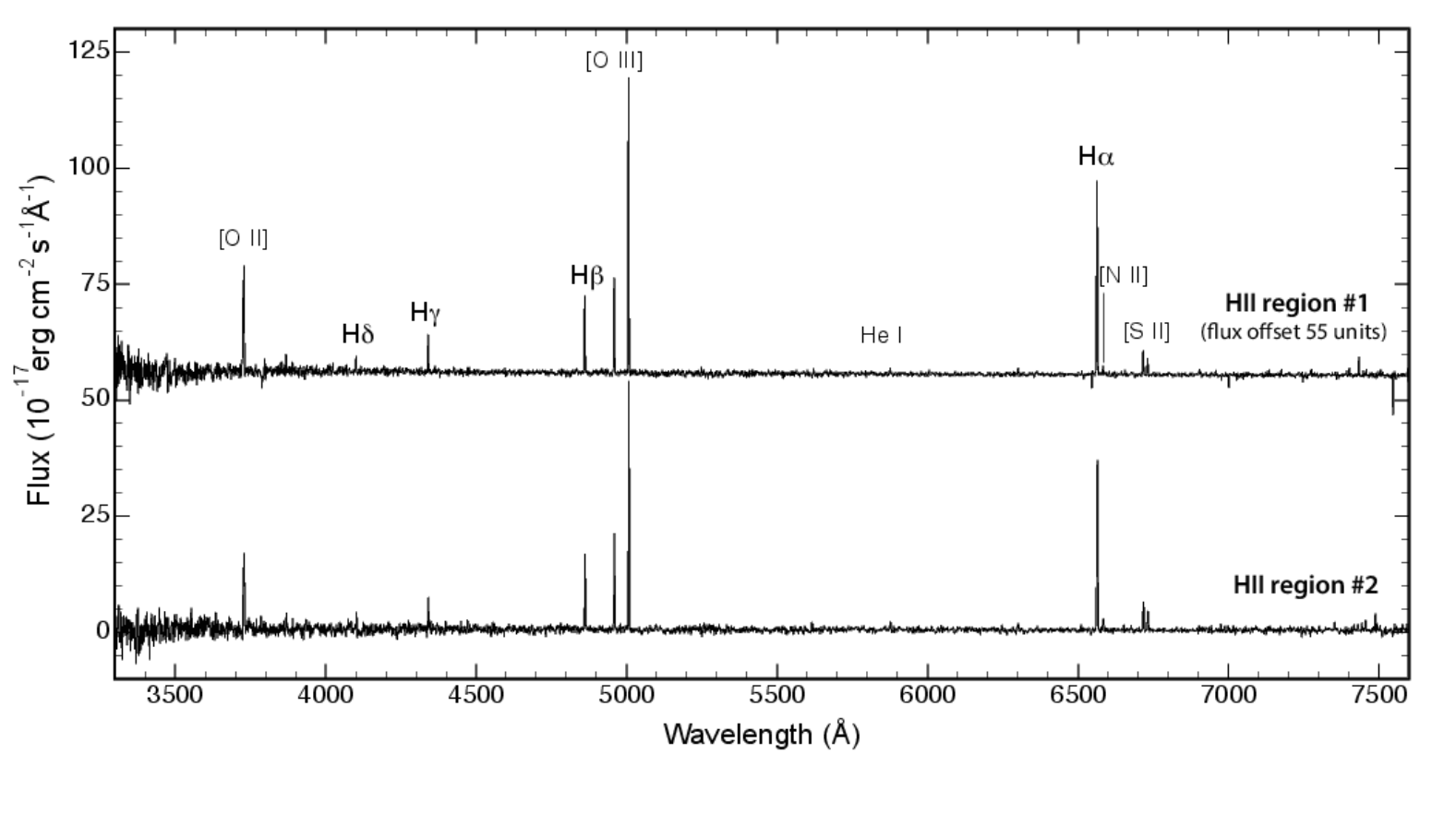}
\end{centering}
\caption{The extracted spectra of the \HII\ region complexes. The weak
  \NII/H$\alpha$ ratio, and the relative strength of \NII\ to
  \SII\ imply that the chemical abundances in these \HII\ regions are
  appreciably sub-solar.}
\label{fig:HII}
\end{figure*}

In Fig.~\ref{fig:AGN_spectra}, the HER (top spectrum) is
characterised by the presence of very high-excitation species such as
[\ion{Ne}{5}] and [\ion{Fe}{7}]. However, the relative ratios of the
\OI, \NII\ and \SII\ lines in the red, and as already noted above the
\NII/H$\alpha$ ratio implies that the chemical abundance is
appreciably sub-solar. As such this spectrum is quite unlike what is
usually seen in the ENLR of local Seyfert-2 galaxies, which appear to
have super-solar metallicities \citep{Davies16}.

The spectrum of the nucleus of the main galaxy is that of a Seyfert-2
with typically strong \OIII\ emission with respect to H$\beta$, and
strong \NII\ emission with respect to H$\alpha$. Also \SII\ and
\OI\ are both fairly strong. A relatively strong continuum of old
stars is also present. 

\begin{figure}
\begin{centering}
\gridline{\fig{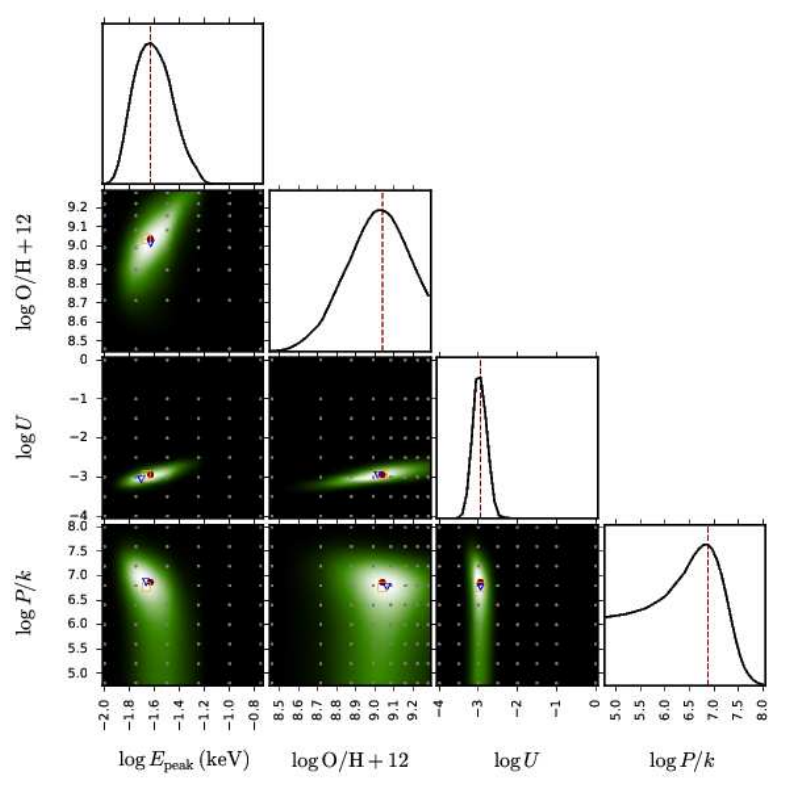}{0.45\textwidth}{AGN
  of ShaSS\,073.}}
\gridline{\fig{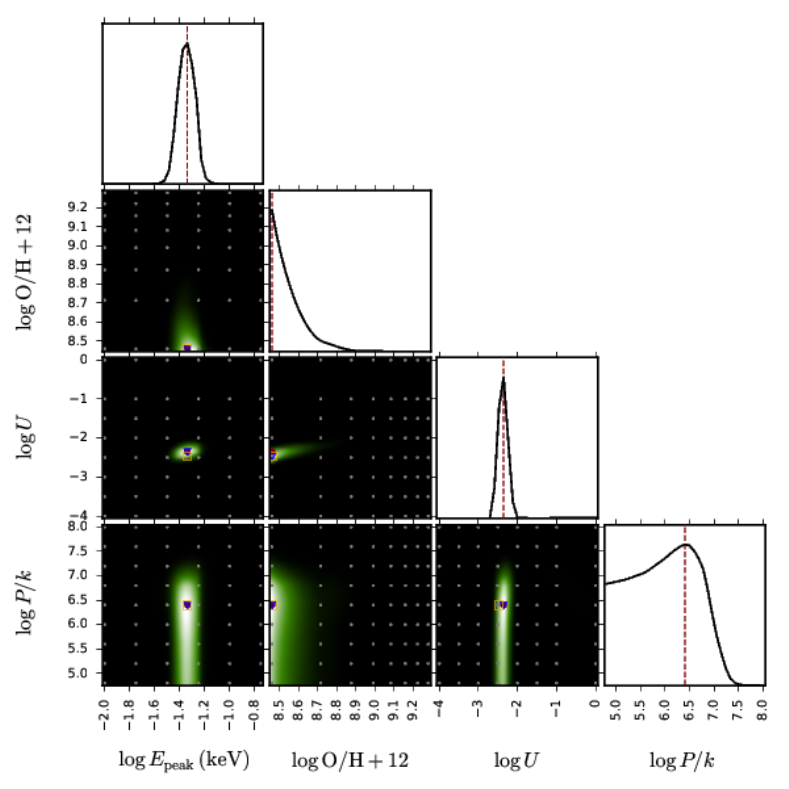}{0.45\textwidth}HER}
\end{centering}
\caption{Marginalised posterior distributions showing constraints on
  the peak energy of the EUV radiation field $E_{peak}$, the oxygen
  abundance $\log \mathrm{O/H} + 12$, the ionization parameter $\log
  U$, and the gas pressure $\log P/k$. Different shades of green (dark
  to light) indicate probability densities from low to high. Gray
  dots represent model grid-points. The 1D marginalized posteriors are
  drawn with black lines, with the brown vertical dotted line at the
  peak of a 1D posterior indicating the value taken as the parameter
  estimate. On each distribution is plotted the model defined by the
  parameter estimates (brown circle), the peak of the 2D marginalized
  distribution (blue open triangle) and the projected peak of the full
  $n$-dimensional PDF (orange open square).}
\label{fig:AGN_bayes}
\end{figure}

\subsection{Bayesian Photoionization Analysis}

As noted above, the theoretical grid of models is at least a 5-D
space: chemical abundances, ionization parameter, gas pressure, dust
depletion factors, and the shape of the EUV spectrum (which may itself
be characterised by more than one variable). In order to objectively
locate the position of an observed object in this space we require a
Bayesian approach to the problem.

In this section we apply the new package NebulaBayes, a generalization
of IZI \citep{Blanc15}, which constrains nebular parameters using a
novel Bayesian approach. Whereas IZI works in two dimensions,
constraining only the ionization parameter and metallicity,
NebulaBayes works in arbitrary dimensions and is parameter
agnostic. The method compares observed line fluxes to an extensive
grid of pre-computed photoionization models, thereby calculating the
relative probabilities of the model parameters at all points in the
multi-parameter space. NebulaBayes will be described in more detail
and made available to the community in the near future (Thomas et al.,
in preparation).

The computed \HII\ region grid has three parameters - the metallicity
$12 + \log \mathrm{O/H}$, the ionization parameter $\log U$, and the
gas pressure $\log P/k$. The logarithmic dust depletion factor is kept
fixed at $-1.5$ for Fe. For the input EUV spectrum of \HII\ regions we
use the spectra described by \citet{Dopita13} which are derived from
the {\tt Starburst 99} code \citep{SB99}.

The NLR photoionization grid has metallicity $12 + \log \mathrm{O/H}$,
ionization parameter $\log U$, and gas pressure $\log P/k$ plus a
fourth parameter, which describes the peak energy of the accretion
disk emission $E_\mathrm{peak}$ (measured in $\log_{10} \,
E/\mathrm{keV}$). This fourth parameter derives from the AGN ionizing
continuum model OXAF \citep{Thomas16} which is used in the MAPPINGS
modelling.

The computational approach is as follows: Firstly, a {\it prior} is
defined over the parameter space to include prior expectations of
parameter values in the Bayesian parameter inference. Secondly, the
{\it likelihood} (the probability of the data given the model,
i.e.\ given each set of parameters in the parameter space) is
calculated assuming Gaussian errors on the observed fluxes and
weighting all emission lines equally. The {\it posterior} (which is
the probability of the model given the data, and describes the
relative probabilities of sets of parameters in the parameter space)
is then calculated according to Bayes Theorem by multiplying the prior
and the likelihood. Parameter estimates and credible intervals are
obtained by marginalizing\footnote{If $N$ is the number of parameters,
  this means to integrate over $N-n$ dimensions to obtain the
  $n$-dimensional distribution for $n$ parameters.} the posterior
probability distribution function (PDF) to obtain 1D PDFs for each
nebular parameter.

\begin{deluxetable}{lcccc}
\tablecaption{NebulaBayes parameter estimates with $68\%$ confidence intervals. \label{tab:NB_results}}
\tablecolumns{5}
\tablewidth{60mm}
\tabletypesize{\footnotesize}
\tablehead{
{\bf Region} & {\bf \HII\ \#1} & {\bf \HII\ \#2} & {\bf Ion. cone} & {\bf AGN} \\}
\startdata
12+$\log$O/H   & 8.4$\pm$0.1       & 8.6$\pm$0.1      & $< 8.5$ &  9.0$\pm$0.2 \\
\hline
$\log(P/k)$    & 6.5$^{+0.1}_{-2.0}$       & 6.5$^{+0.1}_{-1.8}$      & 6.4$^{+0.3}_{-1.2}$      & 6.9$^{+0.3}_{-1.5}$      \\
\hline
$\log U$       & -3.1$\pm$0.1      & -3.1$\pm$0.1     & -2.4$\pm$0.2     & -3.0$^{+0.3}_{-0.2}$     \\
\hline
log($E_{peak}$/keV)  & --        & --       & -1.3$\pm$0.1     & -1.6$\pm$0.2     \\
\enddata
\end{deluxetable}

The [S\,{\sc ii}] $\lambda\lambda 6731/6717$ ratio is the main prior
used, since this directly constrains the gas pressure. For the HER
material, the excitation as derived from the He\,{\sc ii} $\lambda $
4686/He\,{\sc i} $\lambda$ 5876 ratio was also used as a prior. For
these priors, the observed line ratio is compared to a predicted ratio
and a probability is obtained by taking into account the observational
errors on the observed ratio. This calculation at every grid point in
the parameter space yields an $n$-dimensional PDF. The PDFs for each
line ratio prior are multiplied together (weighted equally), and the
resulting prior PDF is normalised before being used in Bayes Theorem.
 
In Fig.~\ref{fig:AGN_bayes}, we show the computed posterior
distributions of the peak energy of the EUV radiation field
$E_{peak}$, $\log \mathrm{O/H} + 12$, the ionization parameter $\log
U$, and the gas pressure $\log P/k$ for the AGN in ShaSS\,073, and the
HER. The dotted lines show the peak of the 1D marginalized
PDF. On each distribution is plotted the model defined by the
parameter estimates (brown circle), the peak of the 2D marginalized
distribution (blue open triangle) and the projected peak of the full
$n$-D PDF (orange open square).

Note that for ShaSS\,622, the oxygen abundance falls below the minimum
value computed in the theoretical grid, so the derived abundance is
only an upper limit. Note also that both $E_{peak}$ and $\log U$ are
appreciably lower in ShaSS\,073 than in the HER. These both
suggest that the AGN in ShaSS\,073 was in a much higher state of
activity when the photons now reaching the HER were launched.

The derived parameter estimates with their likely ranges are presented
in Table~\ref{tab:NB_results} for both the HER and for the two
\HII\ regions. These results confirm that $\log \mathrm{O/H} + 12$ is
much lower in ShaSS\,622 than in ShaSS\,073.

\subsection{Detailed Photoionization Models}
\label{sec:DPM}

In this section we consider another method of exploring how the
physical parameters affect the nebular physics, which involves running
tailored photoionization models for the two \HII\ regions and the HER.

This modelling is performed in a similar manner to that described by
\citet{Dopita14}. This consists in running a fine grid of models,
varying the metallicity and ionization parameter $\log U$ while
keeping the input EUV spectrum, gas pressure and dust depletion
factors fixed. In order to measure the goodness of fit, we measure the
L1-norm, which is the modulus of the mean logarithmic difference in
flux (relative to H$\beta$) between the model and the observations;
\begin{equation}
{\rm L1} ={ \Sigma}_n \left | \log \left[ {F_n({\rm model})} \over {F_n({\rm obs.)}} \right]\right | /n. \label{L1}
\end{equation}
This weights fainter lines equally with stronger lines, and is
therefore more sensitive to the values of the input parameters.
 
\floattable 
\begin{deluxetable}{lc|rr|rrr} 
\tabletypesize{\small}
\tablecaption{The spectrum of the HER
  (Fig.~\ref{fig:AGN_spectra}) and the spectra of two \HII\ regions
  shown in Fig.~\ref{fig:HII} compared with their detailed
  photoionization fits.
\label{tab:fits}} \tablehead{
  \colhead{$\lambda$~(\AA)} & \colhead{Line ID} & \multicolumn{2}{c}{HER} & \multicolumn{3}{c}{\HII\ regions} \\
  \colhead{} & \colhead{} & \colhead{Observed} & \colhead{Model} &
  \colhead{\HII\ \#1} & \colhead{\HII\ \#2} & \colhead{Model} \\}
\startdata
3346 &  [Ne\,{\sc v}]   & $13.19 \pm 0.30$ & 8.39 &&& \\
3426 &  [Ne\,{\sc v}]   & $35.50 \pm 0.41$ & 23.50 &&& \\
3729 & \OII     && & $252.7\pm 6.5$ & $201.5\pm 3.2$ & 355.0 \\
3738 & \OII    & $240.8 \pm 1.6$ & 160.84 &&& \\
3869 & [Ne\,\sc{ iii}]  & $73.07 \pm 1.13$ & 84.67 & $21.8 \pm 3.3$ & $26.0\pm 1.4$  & 52.9 \\
3889 & H$\zeta$   && &  $13.1\pm 1.8$ & $12.8\pm 1.7$  & 10.5  \\
3967 & [Ne\,\sc{ iii}]  & $24.42 \pm 0.63$ & 25.51 &&& \\
3970 &  H$\epsilon$	& $14.82 \pm 1.96$ & 15.40 &&& \\
3889 & H$\zeta$    && &  $13.1\pm 1.8$ & $12.8\pm 1.7$  & 10.5 \\
4068 & \SII & $3.1  \pm 0.53$ & 1.80 &&& \\
4076 & \SII & $1.0  \pm 0.50$ & 0.58 &&& \\
4102 & H$\delta$ & $24.48  \pm 1.60$ & 25.20 & $29.8\pm 1.5$ & $23.8\pm 1.2$  & 25.8 \\
4340 & H$\gamma$ & $45.96  \pm 0.11$ & 46.74 & $48.7\pm 1.9$ & $42.2\pm 1.6$  & 47.0 \\
4363 & \OIII   &  $21.31  \pm 0.17 $ & 20.65 & $8.7\pm 3.4$ &  $0.4\pm 0.4$  & 4.2  \\
4471 & He\,{\sc i}      & $2.80  \pm 0.17$ & 2.81 &&& \\
4686 & He\,{\sc ii}      & $26.4  \pm 0.39$ & 33.58 &&& \\
4711 & [Ar\,{\sc iv}]     & $4.12  \pm 0.13$ & 5.73 &&& \\
4740 & [Ar\,{\sc iv}]    & $3.45  \pm 0.20$ & 4.34 &&& \\
4861 & H$\beta$  & $100  \pm 0.80$ & 100 &  $100.0\pm 2.4$ & $100.0\pm 2.6$ & 100.0 \\
4959 & \OIII   & $292.3  \pm 0.55$ & 331.0 & $116.4\pm 2.4$ & $109.2\pm 2.6$ & 120.2 \\
5007 & \OIII   & $876.3  \pm 0.70$ & 956.7 & $349.1\pm 4.5$ & $327.5\pm 5.4$ & 347.4 \\
5200 & \NI     &  $1.25  \pm 0.60$ & 2.85 &&&  \\
5876 & He\,{\sc i}       & $7.49  \pm 0.10$ & 7.30 & $8.4\pm 1.3$ &  $14.2\pm 2.1$ &  11.5 \\
6087 & [Fe\,{\sc vii}]  & $1.45  \pm 0.26$ & 0.44 &&& \\
6300 & [O\,{\sc i}]     & $14.43  \pm 1.07$ & 12.19 & $10.0\pm 2.0$ &   $7.1\pm 3.9$ &   5.6 \\
6312 & [S\,{\sc iii}]   & $2.89   \pm 0.69$& 8.42 &&& \\
6364 & [O\,{\sc i}]     & $4.56\pm 0.36$ & 3.90 & $6.4\pm 1.6$ &   $3.3\pm 0.9$ &   1.8 \\
6548 & \NII    & $ 7.26  \pm 0.65$ & 5.11 & $4.1\pm 2.3$ &   $7.4\pm 3.5$ &   7.1 \\
6563 & H$\alpha$ & $285.8  \pm 0.30$ & 293.8 & $286.0\pm 4.5$ & $286.0\pm 5.5$ & 285.5 \\
6583 & \NII    & $21.27  \pm 0.96$ & 15.09 & $12.2\pm 3.0$ &  $22.1\pm 3.0$ &  20.2 \\
6678 &  H\,{\sc i}     & $2.13  \pm 0.50$ & 2.06 & $4.9\pm 2.0$ &   $7.0\pm 3.5$ &   3.2 \\
6716 & \SII    & $39.65  \pm0.43$ & 26.58 & $33.9\pm 3.5$ &  $48.9\pm 4.5$ &  23.3 \\
6731 & \SII    & $27.81  \pm 0.35$ & 18.84 & $22.9\pm 2.6$ &  $35.2\pm 3.5$ &  19.7 \\
7136 & [Ar\,{\sc iii}]    & $8.80  \pm 0.61$ & 10.06 &&& \\
7319 & \OII    & $3.54  \pm 0.99$ & 2.50 &&& \\
7329 & \OII    & $2.34  \pm 1.05$ & 2.02 &&& \\
7751 & [Ar\,{\sc iii}]    & $1.89  \pm 0.15$ & 2.41 &&& \\
\enddata
\end{deluxetable}

The results of the fit are shown in Table~\ref{tab:fits}. For the two
\HII\ regions the model has an abundance of 0.3 solar
($12+\log({\mathrm O}/{\mathrm H}) = 8.2$), a pressure $\log(P/k) =
6.5$\,cm$^{-3}$K and ionization parameter $\log U=-2.75$. Together
with the diagnostic diagrams, the Bayesian fitting and this detailed
models we can be confident that in ShaSS\,622 $8.2 \leqslant
12+\log({\mathrm O}/{\mathrm H}) \leqslant 8.5$.
 
In the case of the AGN fitting the hardness of the EUV radiation field
field is strongly constrained by the [\ion{He}{2}]$\,\lambda 4686$ to
[\ion{He}{1}]$\,\lambda 4471$ ratio and the [\ion{He}{1}]$\,\lambda
5876$ to H$\beta$ ratio, since these are largely determined by the
relative number of photons able to ionize H$^0$ to H$^+$, He$^0$ to
He$^+$ and He$^+$ to He$^{++}$. Further constraints on the hardness of
the radiation field are provided by the [\ion{Ne}{5}]\, $\lambda 3426$
to [\ion{Ne}{3}] $\lambda 3868$ ratio and the [\ion{Ar}{4}] $\lambda
4711$ to [\ion{Ar}{3}] $\lambda 7136$ ratio. Our best fit model for
the HER has $\log E_{peak}= -1.25$, a hard photon spectral
index of -1.9, an abundance of 0.4 solar ($12+\log({\mathrm
  O}/{\mathrm H}) = 8.32$), a gas pressure of $\log(P/k) =
6.2$\,cm$^{-3}$K and an ionization parameter $\log U=-2.2$. The
computed L1-norm is 0.11 using 32 emission lines.

A quantity of interest for our analysis is the recombination
time-scale for the gas in the HER. As the gas freely recombines, the
emitted spectrum shifts rapidly to be dominated by the low-ionization
species. [O\,{\sc iii}] therefore rapidly becomes weaker, and
[Ne\,{\sc v}] disappears, while both lines are strong in the spectrum
of HER (Fig.~\ref{fig:AGN_spectra} upper panel). The {\it maximum}
time this process can take is governed by the Hydrogen recombination
time-scale. We assume a Case B recombination coefficient at $10^4$\,K
of $\alpha = 1.58\times 10^{-13}$\,cm$^3$s$^{-1}$ \citep{OF06}. With
the same temperature and the estimated pressure in the HER of
$\log(P/k) = 6.2$\,cm$^{-3}$K, we can infer an Hydrogen density of
$\sim 70$\,cm$^{-3}$ and an electron density of $\sim 80$\,cm$^{-3}$
(allowing for He to be ionised). This leads to a recombination
time-scale of H within the HER of $\sim 2500$\,yr. The recombination
time for [O\,{\sc iii}] ($\alpha = 1.72\times
10^{-11}$\,cm$^3$s$^{-1}$) is $\sim 23$\,yr.
                                                      
\section{Structure of the ShaSS\,622-073 system}
\label{sec:struct}

\begin{figure}
\begin{centering}
\includegraphics[width=80mm]{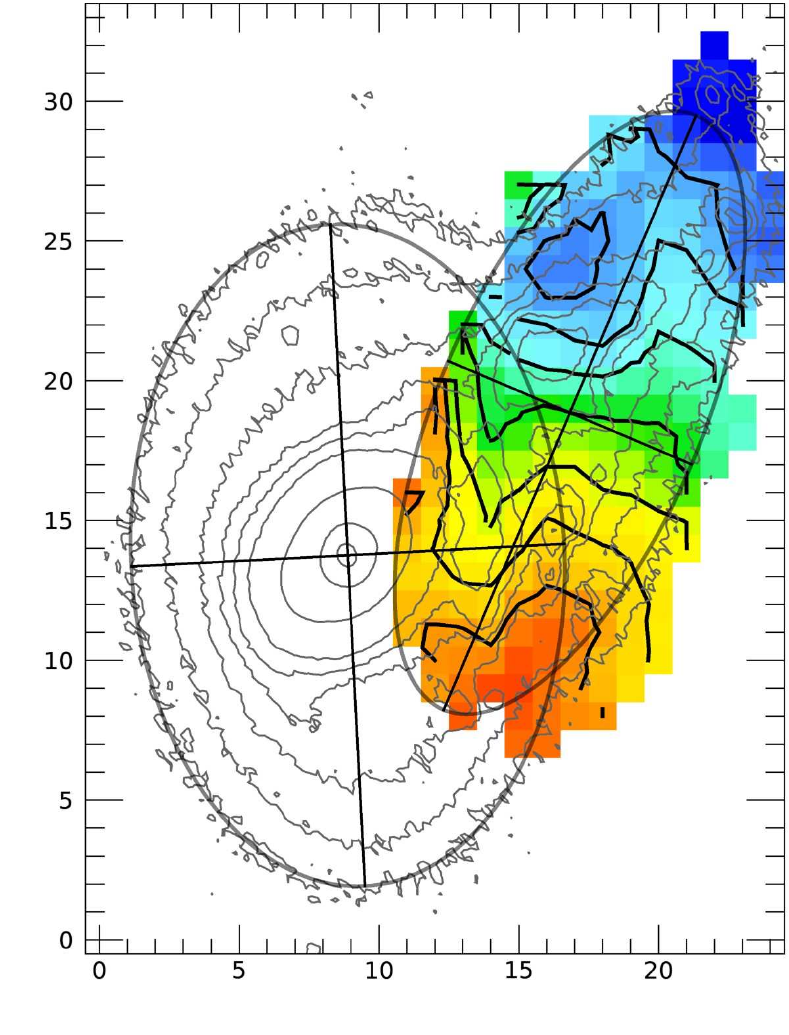}
\end{centering}
\caption{Radial velocity field of the gas of ShaSS\,622.  These are
  the same data shown in Fig.~\ref{fig:kinetic}, but with a different
  choice of the kinematic center aimed at revealing the regularity of
  the velocity field of ShaSS\,622. Ellipses locating the disks of the
  two galaxies are plotted along with the main axes. Black contours of
  constant radial velocity are plotted at 25\,km s$^{-1}$ intervals,
  and gray contours are $r$-band isophotes. Scales are in arcseconds
  in the WiFeS reference frame.}
\label{fig:kin622}
\end{figure}

\begin{figure}
\begin{centering}
\includegraphics[width=80mm]{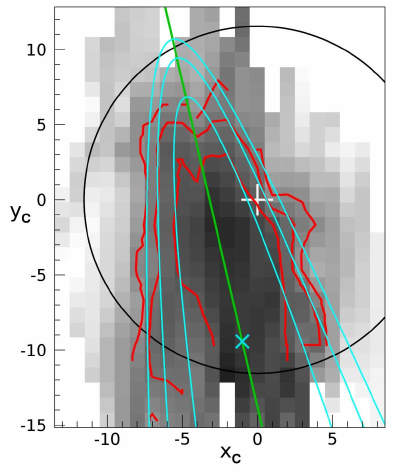}
\end{centering}
\caption{The distribution of [O\,{\sc iii}]/H$\beta$ on the disk of
  ShaSS\,622 (black circle), mapped by gray tones and by the red
  curves of constant ratios log([O\,{\sc iii}]/H$\beta$)=0.85,0.9,0.95
  (outside-in). The green line marks the preferential direction of the
  distribution of line ratios. The white cross is the center of
  ShaSS\,622 and the cyan cross is the assumed intersection of the
  ionization cone axis with the disk. The cyan conics are the
  intersections with the disk of cones of semi-aperture of
  7$^{\circ}$,10$^{\circ}$,12$^{\circ}$(inside-out). The
    outermost conics delimits the HER. The units on the axes are
  arcsecs (measured from the center of ShaSS\,622).}
\label{fig:expanded}
\end{figure}

In this Section we will try to model the three-dimensional structure
of the galaxy ShaSS\,622-073 starting from the observation that the
most of the disk of ShaSS\,622 is kinematically almost unperturbed.
Fig.~\ref{fig:kin622} shows the same radial velocity field of
Fig.~\ref{fig:kinetic}, but assuming a different kinematic center and
removing most of the field of ShaSS\,073 for clarity. An ellipse was
sketched following the external $r$-band isophotes of ShaSS\,073 in
the area where there is no clear superposition between the two
galaxies. Another ellipse encircles the area in which the kinematic
field of ShaSS\,622 appears to be fairly regular, with the center
adjusted to maximize its symmetry. The requirement of maximum symmetry
gave little room for the choice of the center, which is confined
within about $\pm$1 pixel from the chosen position.

The regularity of the velocity field inside the ellipse implies that
the motion of the gas is still only weakly perturbed by the
interaction, in agreement with the very early stage of interaction
claimed for in the preceding Sections\footnote{The lack of axial
  symmetry, apparent from the twisting of iso-velocity contours, might
  be either the initial effect of the interaction or the effect of a
  pre-existing bar causing non-circular motions}. As a consequence of
this, we can safely assume that the gas within the sketched ellipse
still lies in the galactic plane. As we will show in the following,
this has a significant impact on our knowledge of the ShaSS\,622-073
system, because this gas constitutes a nearly flat `projection screen'
being hit by the ionization cone produced by the AGN.

Our analysis starts from the properties of the disks (axis ratios and
position angles) and the distribution of the [O\,{\sc iii}]/H$\beta$
ratio on the disk of ShaSS\,622. The derivations are presented in
Appendix~\ref{AppA}. The axis ratio of ShaSS\,622, $b/a\sim 0.41$,
corresponds to an inclination (disk to sky plane) of
$\pm$66$^{\circ}$(assuming an infinitely thin layer of gas,
appropriate for a late-type spiral). For the inclination we adopt the
notation $k_6\cdot i_6$=$\pm$66$^{\circ}$, where $i_6=66^{\circ}$ and
the parameter $k_6=\pm1$ determines which side of the disk of
ShaSS\,622 is facing the observer and in turn the orientation of the
observer with respect to the whole system (see below).

Let us now introduce a reference system ${\mathcal C}$:($x_{\mathcal
  C}$,$y_{\mathcal C}$,$z_{\mathcal C}$) (see Appendix~\ref{AppA}),
with the origin and the $z_{\mathcal C}$ axis coincident with the
center and the rotation axis of ShaSS\,662 respectively. The
transformation to this system is achieved by a rotation around the
apparent major axis of the galaxy by one of the two angles $k_6
i_6=\pm 66^{\circ}$. In the Appendix~\ref{AppA} it is shown that this
transformation of coordinates can't be applied {\it a priori} to
ShaSS\,073 unless we have an independent constraint on its position.

We will now show that the shape of the illuminated area on
the disk of ShaSS\,622, as defined by the [O\,{\sc iii}]/H$\beta$ ratio, 
allows us to determine the direction to the AGN in the reference
system ${\mathcal C}$, thus providing the needed constraint.

We simulate the transformation to system ${\mathcal C}$ of the disk of
ShaSS\,622 by expanding the observed map of [O\,{\sc iii}]/H$\beta$ by a
factor $1/cos(i_6)$ in the direction orthogonal to the apparent major
axis. This stretch factor transforms the disk of ShaSS\,622 into a
circle and can be shown to correspond exactly to the rotation of $i_6$
degrees for a thin disk\footnote{We performed numerical simulations
  with different shapes and spatial sampling to assess the ability to
  recognize in this way the original shapes on a plane from a skewed
  line of sight. We found that this method recovers shapes quite
  accurately, except for extremely poor spatial sampling,
  which is not our case.}.

Fig.~\ref{fig:expanded} shows the distribution of [O\,{\sc
    iii}]/H$\beta$ on the disk of ShaSS\,622, mapped by gray tones
and by the red curves of constant log([O\,{\sc
    iii}]/H$\beta$)=0.85,0.9,0.95 from outside-in (to compare with the
values in Fig.~\ref{fig:KBPT}). The map and the curves clearly
indicate a preferential direction, marked with a green line, which
should correspond to the projection on the galaxy plane of the axis of
the ionization cone, and therefore also marks the direction to the
AGN. As it is shown in the Appendix~\ref{AppA}, this is the constraint on the
position of ShaSS\,073 which allows us to derive a 3-D model for
the two galaxies.

\begin{figure}
\begin{centering}
\includegraphics[width=80mm]{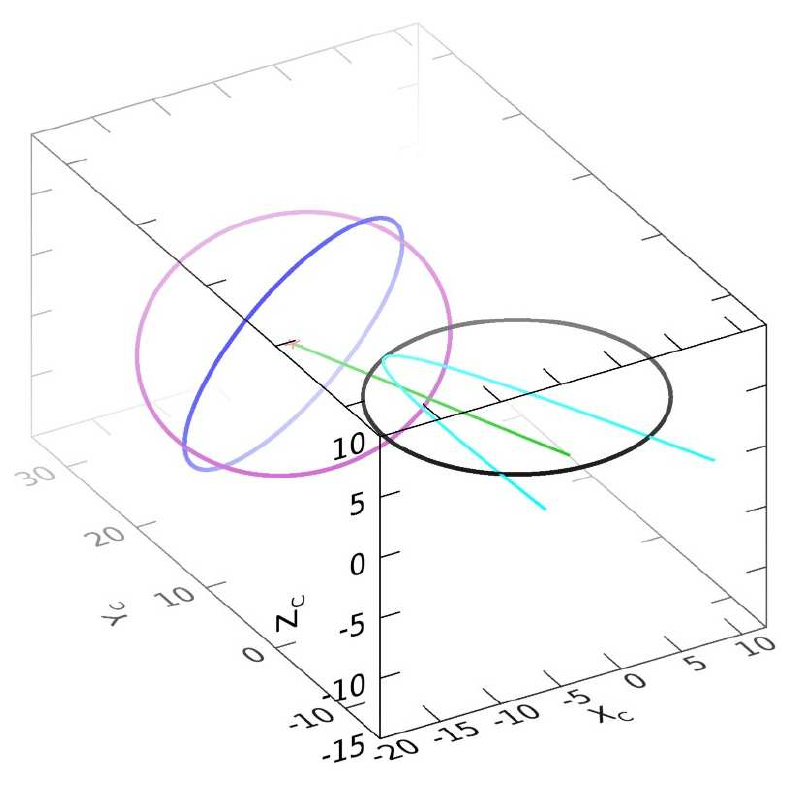}
\end{centering}
\caption{View of the ShaSS\,073-622 pair in reference system
  ${\mathcal C}$, where the axis of ShaSS\,662 coincides with
  $z_{\mathcal C}$ (see Appendix~\ref{AppA}). The disks of ShaSS\,622
  (black) and ShaSS\,073 (blue and purple for $k_7=\pm1$ respectively)
  are shown. Also shown are the axis of the ionization cone (green
  line) and the intersection of the cone of semi-aperture 12$^{\circ}$
  with the disk of ShaSS\,622 (cyan curve, delimiting the HER). The
  units on the axes are kpc.}
\label{fig:3D}
\end{figure}

\floattable
\begin{deluxetable}{llcl}
\tablecaption{Parameters defining the spatial structure of the
  ShaSS\,622-073 system. The intervals of parameters which are broadly
  consistent with the data are indicated in square brackets (see
  Appendix~\ref{AppA}.  \label{tab:3D-geom}} \tablecolumns{4}
\tabletypesize{\footnotesize}
\tablehead{\colhead{} & \colhead{$k_7=-1$} & \colhead{} & \colhead{$k_7=+1$}}
\startdata
\hline
distance between the centers of the galaxies& & 21\,kpc \sl{[19-24]} & \\
\hline
distance along the line of sight & & 19\,kpc \sl{[17-22]} & \\
\hline
angle between their disks & 27$^{\circ}$ \sl{[22-32]} && 69$^{\circ}$ \sl{[63-75]} \\
\hline
semi-aperture of ionization cone &  & 12$^{\circ}$ \sl{[12-15]} & \\
\hline
angle between ShaSS\,073 axis and cone axis & 71$^{\circ}$ \sl{[66-78]} && 28$^{\circ}$ \sl{[21-34]} \\
\hline
angle between cone axis and line of sight && 22$^{\circ}$ \sl{[20-25]} & \\
\hline
angle between the cone axis and the disk of ShaSS\,622 & & 6$^{\circ}$ \sl{[2-10]} &\\
\hline
\enddata
\end{deluxetable}

\begin{figure}
\begin{centering}
\includegraphics[width=80mm]{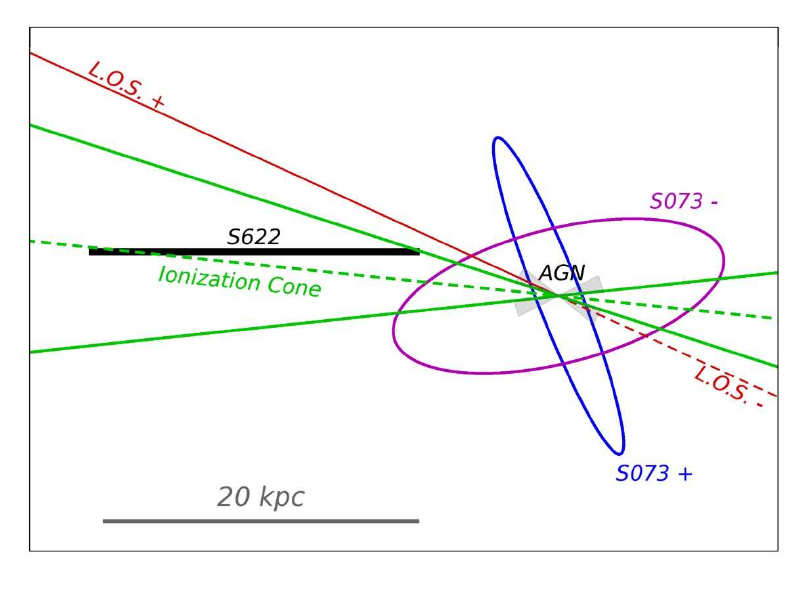}
\end{centering}
\caption{View of the ShaSS\,073-622 system with the axis of the
  ionization cone (green dashed line) in the plane of
  the plot. The disks of ShaSS\,622 and ShaSS\,073 (in the two
  possible orientations according to the sign of $k_7$) are shown,
  with ShaSS\,622 viewed edge-on. The two continuous green lines are
  the walls of the ionization cone, assumed to have a semi-aperture
  $\psi=12^{\circ}$. The two possible line-of-sights are shown as a
  continuous ($k_6=+1$) and dashed ($k_6=-1$) red lines. The gray area
  sketches the angular extent of the outflow which roughly reproduces
  the geometry of the observations.}
\label{fig:cono}
\end{figure}

Knowing now the relative positions of the two galaxies and that the
intersection of the cone axis with the plane of ShaSS\,622 must lie
along the green line in Fig.~\ref{fig:expanded}, we can determine, for
any semi-aperture $\psi$ of the cone, the shape of the area
illuminated by the radiation from the AGN and compare this with the
observed curves of constant [O\,{\sc iii}]/H$_{\beta}$. It turns out
that the range of positions and cone semi-apertures leading to the
observed shape of the illuminated area is relatively small. In other
words, the observed distribution of [O\,{\sc iii}]/H$\beta$ imposes
tight constraints on the geometry (aperture and orientation) of the
ionization cone. The assumed intersection of the cone axis with the plane of
ShaSS\,622 is marked by a cyan cross in Fig.~\ref{fig:expanded}, while
the intersections of cones with semi-apertures of 7$^{\circ}$,
10$^{\circ}$, and 12$^{\circ}$ are shown by the cyan conics. We remark
here that our model gives a purely geometric prediction in terms of
cone-plane intersection, while the actual distribution of [O\,{\sc
    iii}]/H$\beta$ depends on much more complex factors, like, for
instance, the distribution of the nuclear radiation within the cone,
the spatial distribution of dust in the disk and the local physical
conditions of the gas. Taking into account the above caveats, these
curves overlap fairly well with those of constant [O\,{\sc
    iii}]/H$\beta$, and we adopt $\psi=12^{\circ}$ as the
semi-aperture of the ionization cone of the AGN.

The overall picture suggested by our model is depicted in
Figs.~\ref{fig:3D} and \ref{fig:cono} and summarized in
Table~\ref{tab:3D-geom}. Fig.~\ref{fig:3D} shows the galaxies embedded
in the reference system ${\mathcal C}$ from an arbitrary angle of
view. The disk of ShaSS\,622 (black) is affected by the ionization
cone within the area limited by the cyan curve (the HER). The axis of
the ionization cone is plotted as a green line running from the AGN to
the disk of ShaSS\,622. The two possible orientations of the disk of
ShaSS\,073, which are identified by the parameter $k_7$, are shown in
blue ($k_7=+1$) and purple ($k_7=-1$). Fig.~\ref{fig:cono} offers a
view perpendicular to the ionization cone axis (dashed green
line). The surface of the ionization cone is represented by the
continuous green lines, and the galaxies discs are represented as in
Fig.~\ref{fig:3D}, with ShaSS\,622 viewed edge-on. The red lines mark
the two different line of sights according to the sign of $k_6$. Note
that for $k_6=+1$ ShaSS\,622 is closer to the observer than ShaSS\,073
and vice-versa. The gray areas schematically indicate the corona of
the cone where the outflow is taking place.

The distance between the centers of the two galaxies turns out to be
21\,kpc, with their disks being inclined by $27^{\circ}$($69^{\circ}$)
for $k_7=+1$($-1$) with respect to each other. The axis of the
ionization cone is inclined by $22^{\circ}$ with respect to the line
of sight, by $71^{\circ}$($28^{\circ}$) for $k_7=+1$($-1$) with
respect to the rotation axis of the host galaxy, and $6^\circ$ with
respect to the disk of ShaSS\,622. We notice however that the angle
between the cone axis and the disk of ShaSS\,073 for $k_7=-1$ is just
$19^{\circ}$, while we observe the outflow extending beyond
$22^{\circ}$ from the cone axis (see Fig.~\ref{fig:cono} and sixth row
of Table~\ref{tab:3D-geom}). This would imply that the outflow impacts
the disk directly, giving rise to strong kinematical
perturbations. These perturbations should be observable in the gas
kinematics on {\em both} sides of the AGN, since the outflow would be
tangential with the disk. Since no such perturbation is seen, we tend
to conclude that the angle between the ionization cone and the
axis(disk) of the AGN host galaxy is $28^{\circ}$($62^{\circ}$),
corresponding to $k_7=+1$, which also implies that the angle between
the two disks is $69^{\circ}$ (blue models in Figs.~\ref{fig:3D} and
\ref{fig:cono}).

The kinematics of the gas also allows us to constrain the orientation
of the system with respect to the line of sight. In preceding Sections
(Sects.~\ref{sec:kin2comp} and \ref{sec:lrdiag}) we discussed the area
of the northern disk of ShaSS\,073 in which the gas is characterised
by a second kinematical component, which is blue-shifted with respect
to the disk and with somewhat higher velocity dispersion, but which
does not belong to the outflow. The natural explanation for this
kinematics is the tidal interaction with ShaSS\,622, which is starting
to pull the gas out of the disk of ShaSS\,073 (blue-shift) inducing at
the same time turbulence (high $\sigma$). Looking at
Fig.~\ref{fig:cono}, this can be the case if and only if we are
observing the system from the `upper left' ($k_6 =+1$), otherwise the
perturbed gas would be red-shifted. We therefore tentatively ascribe
the second kinematical component of the gas to the onset of the tidal
perturbation in the disk of ShaSS\,073. In addition, being this
component blue-shifted, we infer that the system is oriented as
parametrized by $k_6=+1$ (Fig.~\ref{fig:cono}), with ShaSS\,622 closer
to us (by $\sim$19\,kpc) than the AGN-host galaxy. 

To assess the reliability of this picture, we changed, within
reasonable limits, the initial assumptions concerning the position and
orientation of the two disks and derived from these new values the
parameters of the 3-D structure. The results, given in square brackets
in Table~\ref{tab:3D-geom}, represent extreme cases which are still
marginally consistent with the data. This process is detailed in the
Appendix~\ref{AppA}. The values in Table~\ref{tab:3D-geom} show that
our description of the spatial structure of the ShaSS\,073-622 system
is robust against reasonable changes of the initial assumptions.

\section{Discussion}
\label{sec:dis}

\subsection{Allocating the AGN radiation power}

An interesting question is whether the flux from the AGN is consistent
with the outflow being powered by radiation pressure and, more
important, can ionize the gas in ShaSS\,622 at the observed level.

An upper limit to the power needed to maintain the outflow may be
obtained starting from its high ionization parameter ($\log U \sim
-1.0$, see Sect.~\ref{sec:models}). Such a high value indicates that
the gas pressure is determined by the radiation pressure in this
component. Furthermore, for this component, the [S II] $\lambda\lambda
6731/6717$ ratio is $\sim 1.3$.  Taking into account the errors, this
indicates an electron density $2.8 \leqslant n_e \leqslant
3.5$\,cm$^{-3}$. When the gas is radiation-pressure dominated, we can
set the gas pressure $P_{gas} = L/4\pi R^2 c $, where $L$ is the
luminosity of the source, and $R$ is the distance of the gas parcel
from the source. For ShaSS\,073, the emitting region is unresolved
within the central spaxel, so we can only compute an upper limit on
the luminosity, assuming that the emitting region fills the
spaxel. This implies $R\sim 500$\,pc and hence $\log L \lesssim 45.0
$\,erg\,s$^{-1}$, which is actually larger than the bolometric
luminosity of the AGN $\log L = 44.39$\,erg\,s$^{-1}$ derived in
Sec.~\ref{sec:SED}.

The \OIII\ luminosity of the gas in the HER can be used to
estimate the AGN luminosity. The integrated \OIII\ $\lambda 5007$ flux
in this region is $2.25\times10^{42}$\,erg\,s$^{-1}$. Using the
geometric model of Sec.~\ref{sec:struct}, we estimate that the ionized
region intercepts a solid angle $\Omega =0.0686$\,sr as seen by the
AGN in the main galaxy. It follows that the total omnidirectional
\OIII\ flux would be $4.12\times10^{44}$erg\,s$^{-1}$. In the
photoionization model presented above, the \OIII\ line represents
8.87\% of the bolometric EUV flux. Therefore, we can estimate the
luminosity of the AGN as $\log L =45.66$\,erg\,s$^{-1}$.
 
Combining our photoionization model estimate with the bolometric
luminosity given by the SED fit, $\log L = 44.39$\,erg\,s$^{-1}$, we
conclude that the AGN radiation required to excite the gas inside the
HER is $\sim 20$ times the current luminosity of the AGN. This is also
supported by the Bayesian analysis which obtains lower values of both
$E_{peak}$ and $\log U$ for the AGN than in the HER. Taking into
account the maximum semi-aperture of the cone allowed by the data
($15^\circ$, see Appendix~\ref{AppA}), we obtain an upper limit for
the solid angle of $\Omega = 0.1070$, which would reduce the above
factor from 20 to 13.

Considering the relative positions of the AGN and ShaSS\,622
(Fig.\ref{fig:cono}) and the fact that the ionized gas recombination
times are very short compared to the light travel time
(Sect.~\ref{sec:DPM}), we infer that the AGN luminosity is dropped by
a factor 20 within the last $\sim 3\times 10^4$\,yr\footnote{This
  value refers to the gas in the HER closest to the AGN, which
  will be the first affected by the fading of the AGN.}.

The bolometric luminosity of the AGN has been estimated in
Sect.~\ref{sec:SED} from the emission at 5--6\,$\mu$m which comes from
the optically-thick dust torus surrounding the central engine and
accretion disk. It is therefore interesting to compare the above
time-scale with the time needed to turn this mid-infrared emission
off. The temperature of the dust and resulting infrared emission
responds rapidly to changes in the accretion rate, as confirmed by
quasars/AGN being variable in the mid-infrared on 1--10\,year
time-scales \citep{KKA16}, and the tight correlation between nuclear
mid-infrared and X-ray luminosities of AGNs
\citep[e.g.][]{LMS04,MCA15}. In the case of an AGN being
instantaneously quenched, the dust torus should cool from 1000\,K to
several tens of Kelvin and the 5--6\,$\mu$m emission drops by a factor
$\sim 300$ within $\sim 30$\,yr \citep{IT17}.

\subsection{Relation of ShaSS\,073-622 with light echoes}

Assuming that our picture of the spatial structure of the system is
correct, and also knowing the relative velocity of the galaxies along
the line of sight (130\,\kss), there is still one major unknown
preventing us to envisage the future of our system - the spatial
relative velocity.

Depending on this velocity, the interaction could result from a
complete merging of the two galaxies to just a pass-by, with the gas
probably being stripped off ShaSS\,622 and captured by the AGN host.
In any case, it is likely that the gas of ShaSS\,622 will cross the
ionization cone of the AGN, while in the meantime it is likewise
possible that the AGN activity will change in intensity.

The ShaSS\,073-622 system can be related to the {\it green bean} (GB)
sample of \citet{SDH13} at redshifts $z=0.2-0.6$ characterized by
strong \OIII\ fluxes. Likewise this sample, the ShaSS\,073-622 system
is characterized by the presence of an AGN and an EELR powered by it
and the GBs are Seyfert-2 galaxies as ShaSS\,073.

The luminosities at 24\,$\mu$m and in the \OIII$\lambda$5007 line of
our system are also consistent with the sample of `quasar light-echo'
candidates of \citet{SDH13}. We remark however that the fact that the
AGN flux is currently in a faded phase is to be considered incidental,
because this variation time scale is much shorter (by a factor
$\sim10^{3-4}$) than the interaction/merging time scale.

Another feature in common with most of light echo objects is the fact
that our AGN is radio-weak. This also holds for several objects of the
Schirmer et al.'s sample as well as for Hanny's Voorwerp
\citep[HV,][]{RGJ10}, which differently from the GBs consists of an
ionized nebula detached from the galaxy IC\,2497 containing the
powering AGN. \citet{KCB12} assembled a sample HV-like galaxies from
the Galaxy Zoo Survey (hereafter GZ sample, see also
Sect.~\ref{sec:intro}) which, as ShaSS\,073-622, are lower in redshift
than the GBs. The diagnostic diagrams in Fig.~\ref{fig:KBPT} can be
compared to the line ratios of the GZ sample \citep[see Fig.~4
of][]{KCB12} and the line ratios of the GB SDSS\,J224024.1-092748
which has been studied in detail by \citet{DST15}. The gas in the
HER populates the diagrams where line ratios typical of
Seyfert galaxies are expected in agreement with the finding of the
other works. In particular, the value of log(\OIII/\Hb)\,$\sim 1$ is
completely consistent with those measured for the mentioned samples.

Previous studies invoke the role of the interaction to explain their
observations. In the GB SDSS\,J224024.1-092748 \citep{DST15}, the EELR
is interpreted as the remnant of a quasar-driven outflow triggered by
a preceding galaxy interaction. Most (73\%) of the objects in the GZ
sample are classified as interacting, merging or post-merging systems,
and \citet{KMB15} identified in eight GZ galaxies tidal tails, shells
and chaotic dust structures and systematically lower metallicity of
the ionized gas with respect to the nuclei. This supports the
hypothesis that the the EELRs around local AGNs are illuminated tidal
debris and ShaSS\,073-622 seems to be the likely progenitor of a GB-
or HV-like object.

\noindent
\section{Summary and conclusions}
\label{sec:CONC}

In the framework of our project aimed at the study of galaxies in the
process of being transformed in the Shapley supercluster by using the
multi-band ShaSS data-set \citep{ShaSSI}, we targeted a system
consisting in a galaxy hosting a Seyfert-2 nucleus (ShaSS\,073) and
another galaxy (ShaSS\,622), which are in the initial phase of
interaction. This study is based on integral-field optical
spectroscopy as well as on photometry from the FUV to the IR and
radio.

The AGN-host galaxy is the more massive of the pair
(M*=5.7$\times$10$^{10}$\,M$_{\odot}$) and is probably of
morphological type (R)SB0a, while the mass of the companion (Sbc) is a
factor $\sim$10 lower.  The available fluxes from FUV (GALEX) through
optical and IR (IRAS, Spitzer) allow to recover the spectral energy
distribution of the nucleus of ShaSS\,073, which shows both the
contribution of the AGN and star formation. The radio luminosity of
L$_{1.4 GHz}\sim2\times 10^{22}$\,W\,Hz$^{-1}$ most probably comes
from star formation at a rate of SFR$\sim6$\,M$_\odot$yr$^{-1}$. Both
the SED and the NIR-MIR fluxes give for the AGN a bolometric
luminosity of $2.4\times 10^{44}$\,erg\,s$^{-1}$.

The chemical abundances of the gas phase are significantly different
in the two galaxies, being $\sim$1.5 super-solar in the AGN host
galaxy and $\sim$0.4 solar in the companion. The dust content in the
two galaxies also reflects this difference. This marked difference is
crucial in recognising the gas from the two galaxies all over the
system.

Clear signs of interaction between the two galaxies are i) the
perturbation of the kinematics of the portion of the disk of
ShaSS\,622 closest to ShaSS\,073; ii) gas from the disk of ShaSS\,073
is pulled toward the companion and iii) the distortion of the external
ring of ShaSS\,073. However, the perturbation is still weak in the
sense that most of the original rotational motion of the two disks is
in place and the two gas populations appear to be still well
separated. Moreover, the estimated distance of 21\,kpc between the two
galaxy centers is consistent with this being their initial approach.

The most striking feature of the ShaSS\,073-622 system is the high
excitation of the gas over a large area ($\sim 170$\,kpc$^2$) of the
disk of ShaSS\,622. The gas is excited by the AGN, as is demonstrated
by different kinds of line-ratio diagnostics. The rest of the gas in
ShaSS\,622 consists of normal \HII\ regions. Around the center of
ShaSS\,073, high-velocity ($v_R\sim $440\,\kss\ $\sigma\sim$490\,\kss)
gas reveals the presence of an AGN outflow extending out to $\sim
2-3$\,kpc and ejecting the dust from the nuclear region, as shown by
the high values of the attenuation around the galaxy center. The
excitation of the external gas and the outflow demonstrate the
presence of the hollow bi-cone structure common in (type 2) AGNs.

The analysis of emission-line fluxes in the different areas of the
system reveals, in addition to the differences in metallicity, the
different status of ionization in i) the AGN ($\log U\sim -3$); ii) in
the HER ($\log U \sim -2.4$); iii) the \HII\ regions (log$U\sim
-3$). The value of the ionization parameter of the AGN is lower than
in the HER, and the same is true for the peak energy of the
radiation. This is contrary to the expectations, considering that the
HER should be powered by the AGN, and this is further motivation
to investigate the radiation energy balance of the system.

The power needed to excite the gas in the HER is
4.6$\times10^{45}$\,erg\,s$^{-1}$, which is a factor 20 larger than
the AGN bolometric luminosity.

The natural explanation of this discrepancy is that the AGN flux has
faded in the time the radiation needs to travel from the nucleus of
ShaSS\,073 to the disk of ShaSS\,622, which is $\sim 3\times
10^4$\,yr, which we therefore assume as an upper limit of the
variability time scale experienced by the AGN. Similar time-scales are
obtained by \citet{KMB15} ($\sim 5 \times 10^4$\,yr) and \citet{KLM17}
($\sim 2 \times 10^4$\,yr).

It is of course impossible to predict whether the ShaSS\,073-622
system will turn into the kind of objects belonging to the family of
GBs or HV-like objects, but it appears that all ingredients are
already in place. Because of these features, the ShaSS\,073-622 system
could well constitute the first example of a parent of HV-like
objects. Considering the fact that both the GB and the HV-like objects
are rare, our discovery suggests that either we caught an almost
unique event or such phenomenon has a short time-scale and is not
intrinsically `dramatic'. Since we do not trust the fortune so much,
we tend to believe that similar events are less rare than previously
believed and we expect that current (e.g. MANGA) and future surveys
with IFS will deliver other such systems improving the knowledge of
this complex phenomenon.

\acknowledgments 

We thank the anonymous referee for her/his comments which helped to
improve this work. This work is based on data collected with (i)
WiFeS at the 2.3 m telescope of the Australian National University at
Siding Spring (Australia) and (ii) OmegaCAM at the ESO INAF–VLT Survey
Telescope and VIRCAM at VISTA, both at the European Southern
Observatory, Chile (ESO Programmes 088.A-4008, 090.A-0094, 090.B-0414,
093.A-0465. The optical imaging is collected at the ESO VLT Survey
Telescope using the Italian INAF Guaranteed Time of Observations. This
publication makes use of data products from i) WISE, which is a joint
project of the University of California, Los Angeles, and the Jet
Propulsion Laboratory/California Institute of Technology, funded by
the National Aeronautics and Space Administration and ii) GALEX
(Galaxy Evolution Explorer) is a NASA Small Explorer launched in April
2003. We gratefully acknowledge NASA’s support for construction,
operation and science analysis for the GALEX mission, developed in
cooperation with the Center National d’Etudes Spatiales of France and
the Korean Ministry of Science and Technology. P. Merluzzi and
G. Busarello acknowledge financial support from PRIN-INAF2014: {\it
  Galaxy Evolution from Cluster Cores to Filaments} (PI
B.M. Poggianti). M. A. Dopita acknowledges the support of the
Australian Research Council (ARC) through Discovery project
DP16010363. This research is supported by an Australian Government
Research Training Program (RTP) Scholarship.

\software{PyWiFeS \citep{Childress14}, LZIFU \citep{HMG16}, Mappings
  5.0 (Sutherland et al. 2017, in prep.), Starburst 99 \citep{SB99},
  IZI \citep{Blanc15}, NebulaBayes (Thomas et al. in prep.)}

\newpage
\appendix

\section{Geometry of the system}
\label{AppA}

The transformation from the reference system of the WiFeS frame
(system $\mathcal{A}$) to system $\mathcal{B}$, where ShaSS\,622 is
centered at the origin, with the major axis along $x_{\mathcal{B}}$ is given by:

\begin{equation}
\begin{array}{lll}
x_{\mathcal{B}} & = &  (x_{\mathcal{A}}-x_{{\mathcal A},6})\cdot cos(\alpha)+(y_{{\mathcal A}}-y_{{\mathcal A},6}) \cdot sin(\alpha)\\
y_{\mathcal{B}} & = & -(x_{\mathcal{A}}-x_{{\mathcal A},6}) \cdot
                      sin(\alpha)+(y_{{\mathcal A}}-y_{{\mathcal
                      A},6}) \cdot cos(\alpha)\\
z_{\mathcal{B}} & = &  z_{\mathcal{A}}
\end{array}
\end{equation}

\noindent where $(x_{{\mathcal A},6}, y_{{\mathcal A}, 6})$ are the
coordinates of the center of ShaSS\,622 in system ${\mathcal A}$ and
$\alpha=67^{\circ}$ is the angle between the major axis of the galaxy
and ${\bf x}_{\mathcal A}$ . The coordinate axis $z_{\mathcal{B}}$ corresponds to
the line of sight.

A subsequent transformation to system ${\mathcal C}$ brings the disk
of ShaSS\,622 on the coordinate plane $z_{\mathcal{C}}$=0:
\begin{equation}
\begin{array}{lll}
x_{\mathcal{C}} & = &  x_{\mathcal{B}} \\
y_{\mathcal{C}} & = &  y_{\mathcal{B}} \cdot cos(k_6
                      i_6)-z_{\mathcal{B}} \cdot sin(k_6 i_6)\\
z_{\mathcal{C}} & = &  y_{\mathcal{B}} \cdot sin(k_6 i_6)+z_{\mathcal{B}} \cdot cos(k_6 i_6)
\end{array}
\end{equation}

where $k_6 i_6=\pm 66^{\circ}$. In these coordinates the rotation axis
of ShaSS\,622 is along $z_{\mathcal{C}}$. We set $k_6=1$ when the
farther side of the disk is on the East in the sky.

Let us denote with the subscript $A$ (for `AGN') the position of the
center of ShaSS\,073 and rewrite the above equations as
\begin{equation} \label{eq:a3}
\begin{array}{lll}
x_{{\mathcal C},A} & = &  x_{{\mathcal B},A} \\
y_{{\mathcal C},A} & = &  y_{{\mathcal B},A} \cdot cos(k_6
                         i_6)-z_{{\mathcal B},A} \cdot sin(k_6 i_6) \\
z_{{\mathcal C},A} & = &  y_{{\mathcal B},A} \cdot sin(k_6 i_6)+z_{{\mathcal B},A} \cdot cos(k_6 i_6) \ \ .
\end{array}
\end{equation}

This is a system of three equations with six variables of which only
two are known: $x_{\mathcal{B},A}$ and $y_{\mathcal{B},A}$. To solve
the system for the coordinates of the AGN in system ${\mathcal C}$ we
need another constraint between the variables. Such a constraint is the
projection of the axis of the ionization cone in the plane
$z_{\mathcal C}$=0, which relates the coordinates of the AGN in that plane.
The equation of this projection is
\begin{equation} \label{eq:a4}
y_{\mathcal C} = P x_{\mathcal C} + Q \ \ ,
\end{equation}
with $P$ and $Q$ determined from the green line in Fig.~\ref{fig:expanded}.
This equation reduces to three the number of unknown variables in the above
system~\ref{eq:a3}, making it possible to derive the full spatial position of
the AGN.
For instance, eqs.~\ref{eq:a3} and~\ref{eq:a4} lead to:
$$z_{{\mathcal B},A}=\frac{y_{{\mathcal B},A} \cdot cos(k_6 i_6)-P \cdot x_{{\mathcal B},A} -Q }{sin(k_6 i_6)} \ \ ,$$
and eventually to the position of the AGN in system ${\mathcal C}$. 

It is now possible to determine the distance $D$ between the centers
of ShaSS\,622 and ShaSS\,073
$$D = (x_{{\mathcal C},A}^2+y_{{\mathcal C},A}^2+z_{{\mathcal
    C},A}^2)^{\frac{1}{2}} \ \ ,$$
since the center of ShaSS\,622 is in the origin.

To derive the relative orientation of the two disks we compute the
angle between the axis of ShaSS\,073 and $z_{\mathcal{C}}$, which is the
axis of ShaSS\,622. The direction vector of the axis of ShaSS\,073 in system
${\mathcal B}$ is given by
$${\bf G}=sin(k_7 i_7)\cdot sin(\phi)\,{\bf x_{\mathcal B}} - sin(k_7
i_7)\cdot cos(\phi)\,{\bf y_{\mathcal B}}+cos(k_7 i_7)\,{\bf
  z_{\mathcal B}} \ \ , $$ where $i_7$=50$^{\circ}$ is the inclination
derived from the axis ratio and assuming a disk of finite thickness
0.13 appropriate for an early-type disk. We introduce the parameter
$k_7=\pm 1$, with $k_7=+1$ corresponding to the case where the eastern
side disk in the sky is nearer to the observer. The angle $\phi$ is
the angle of the major axis with $x_{\mathcal B}$, and $({\bf
  x_{\mathcal B}},{\bf y_{\mathcal B}},{\bf z_{\mathcal B}})$ are the
direction vectors of system ${\mathcal{B}}$.

The angle $\gamma$ between ${\bf G}$ and the axis $z_{\mathcal C}$ is
given by the scalar product $cos(\gamma) = {\bf G} \cdot {\bf z_{\mathcal C}}$,
which gives
\begin{equation}
cos(\gamma)= cos(k_7 i_7)\cdot cos(k_6 i_6) - sin(k_7 i_7)\cdot
cos(\phi)\cdot sin(k_6 i_6) \ \ .
\end{equation}

For the study of the geometry of the AGN ionization cone and the
interaction with the disc of the companion galaxy, let us start
considering a circular cone of semi-aperture $\psi$ with its axis
along $w$ in a reference system ($u$,$v$,$w$), and with the vertex at
(0,0,-$h$). The equation of this cone is
$$u^2+v^2=tan(\psi)^2\cdot(h+w)^2 \ \ .$$
The intersection of the cone with the plane $w=0$ is a circle of
radius $h\cdot tan(\psi)$. 
Let us consider another reference system ($x$,$y$,$z$) with the same
origin, and the $y$ axis coincident with the $v$ axis,
but rotated counter-clockwise by an angle $\xi$ around $v$. 
In this system the equation of the cone becomes
$$(x\cdot cos(\xi)-z\cdot sin(\xi))^2+y^2=tg(\psi)^2\cdot (x\cdot sin(\xi)+z\cdot cos(\xi)+h) \ \ .$$
This is the equation of a cone of semi-aperture $\psi$ whose axis lies
on the $y=0$ plane and is inclined by an angle $\xi$ (positive if
measured clockwise) with respect to the $z$ axis. The vertex of the cone lies 
in the $z<0$ semi-space at a distance $h$ from the origin. 
The intersection of the cone with the plane $z=0$ becomes
$$D\cdot x^2 -2\cdot tan(\psi)^2\cdot h\cdot sin(\xi)\cdot x+ y^2 =
tan(\psi)^2\cdot h^2 \ \ ,$$
where
$$D=cos(\xi)^2 - tan(\psi)^2\cdot sin(\xi)^2 \ \ .$$
The sign of $D$ determines which type of conics is the intersection of
the cone with the plane. 
Since by construction both $\psi$ and $\xi$ belong to the interval
$[0,\frac{\pi}{2}]$, the conditions $D \geq 0$
implies$\frac{\pi}{2} \geq \psi + \xi$.
Note that $D=0$, i.e. $\psi + \xi = \pi/2$, implies that the $x$-axis
is parallel to the surface of the cone, so that the intersection is
the parabola
$$x = \frac{y^2}{2\cdot tan(\psi)^2\cdot h\cdot sin(\xi)}-\frac{h}{2\cdot sin(\xi)} \ \ .$$

If $D\neq 0$, after some algebra, we obtain
\begin{equation} \label{eq:a5}
\frac{(x-x_o)^2}{A^2}+ sgn(D)\cdot \frac{y^2}{B^2}=1 \ \, 
\end{equation}
where
\begin{equation}
\begin{array}{lll}
A &=& h\cdot tan(\psi)\cdot cos(\xi) / |D|  \nonumber \\
B &=&A\cdot |D|^{\frac{1}{2}}  \nonumber \\ .
\end{array}
\end{equation}
This is the equation of an ellipse or an hyperbola depending on
whether $D$ is greater or smaller than zero. From the above relations,
the eccentricity $e$ of the conic is simply given by $e^2 = 1-D $
($e<$1 for an ellipse and $e>$1 for an hyperbola).  Notice that by
construction the origin of the coordinates is the intersection of the
axis of the ionization cone with the plane of ShaSS\,622, while the
center of the conic is $x_o$ (Eq.~\ref{eq:a5}). The distance between
the intersection of the cone axis and the focus of the conics is given
by $x_0-sgn(D)\cdot A\cdot e$ for the ellipse and hyperbola and
$A_{par}-h / 2 sin(\xi)$, where $A_{par} = \frac{1}{2}\cdot h\cdot
tan(\psi)^2\cdot sin(\xi)$ is the distance from the focus to the
vertex of the parabola. The relations derived for the ionization cone,
combined with the preceding ones, allow to determine the other
quantities present in Table~\ref{tab:3D-geom}.

To asses the reliability of the parameters defining the
three-dimensional structure of the system (Table~\ref{tab:3D-geom}),
we repeated the above derivations with different sets of initial
parameters for the two disks - position, ellipticity and
orientation. We considered {\it acceptable} values of the parameters
for ShaSS\,622 those for which the gas velocity field remains fairly
symmetric with respect to the center of the galaxy. For ShaSS\,073
these parameters are well constrained by the unambiguous definition of
the center and the regular (elliptical) shape of the external
isophotes (in the unperturbed region). In any case, we kept the widest
possible ranges for all of the parameters in order to stress their
influence on the results. The adopted ranges of the initial parameters
are listed in Table~\ref{tab:sim}, while Fig.~\ref{fig:sim} shows the
ellipses generated by the parameters within those ranges.

The derived ranges of the parameters defining the spatial structure of
the system are given in square brackets in Table~\ref{tab:3D-geom}.

We note that changing the initial values of the ellipses leads to a
7-dimensional parameter space, which is impossible to keep under
control. Furthermore, many parameters are correlated\footnote{As an
  example, the angle $\zeta$ between ShaSS\,073 and the cone axis and
  the angle $\gamma$ between the disks of the two galaxies are
  anti-correlated, so that there is no system with
  $\zeta$=66$^{\circ}$ and $\gamma$=22$^{\circ}$, because the minimum
  $\gamma$ for these values of $\zeta$ is 28$^{\circ}$.}. The limits
in Table~\ref{tab:3D-geom} are instead given as they were
independent. A more rigorous treatment is however out of our scope,
which is to verify that the values provided by our model are
reasonable and robust against reasonable changes of the assumed
initial parameters.

\floattable 
\begin{deluxetable}{l|cc|cc|c} 
\tabletypesize{\small} \tablecaption{Ranges of the initial parameters
  adopted in order to estimate the reliability of the model. Values
  within the ranges generate the ellipses shown in
  Fig.~\ref{fig:sim}. The subscript `0' indicates the quantities of
  the model adopted in Sec.~\ref{sec:struct}. Units are kiloparsecs
  and degrees. \label{tab:sim}} \tablehead{\colhead{} &
  \multicolumn{2}{c}{Position of the center} & \multicolumn{2}{c}{Main
    axes} & \colhead{Major axis P.A.} \\}
\startdata
ShaSS\,622 & x$_0$-1, x$_0$+1 & y$_0$ & a$_0$, a$_0$+2 &b$_0$, b$_0$+1 & PA$_0$-3, PA$_0$ \\
ShaSS\,073 & x$_0$ & y$_0$ & a$_0$-0.5, a$_0$+0.8 &b$_0$-0.5, b$_0$+0.5 & PA$_0$-3, PA$_0$+1 \\
\enddata
\end{deluxetable}

The range of cone semi-apertures was derived as follows. For each of
the above models, we produced the analogous of
Fig.~\ref{fig:expanded}, and changed the semi-aperture and the
intersection of the cone axis with ShaSS\,622 (cyan cross in
Fig.~\ref{fig:expanded}) to achieve the best match of the theoretical
conics with the curves of constant ratios [O\,{\sc iii}]/H$\beta$ (red
isophotes in Fig.~\ref{fig:expanded}). The vast majority of models
return values close to 12$^{\circ}$, no model gives lower values than
this, while the value of 15$^{\circ}$ is produced in just one
model. In this case the intersection of the cone axis with the disk of
ShaSS\,622 (cyan cross in Fig.~\ref{fig:expanded}) moves to
y$_\mathrm{c} < 20$, well outside of the disk.

This analysis leads us to conclude that the model represented by
Figs.~\ref{fig:3D} and \ref{fig:cono} is a realistic description of
the spatial structure of the ShaSS\,073-622 system.

\begin{figure}
\begin{centering}
\includegraphics[width=78mm]{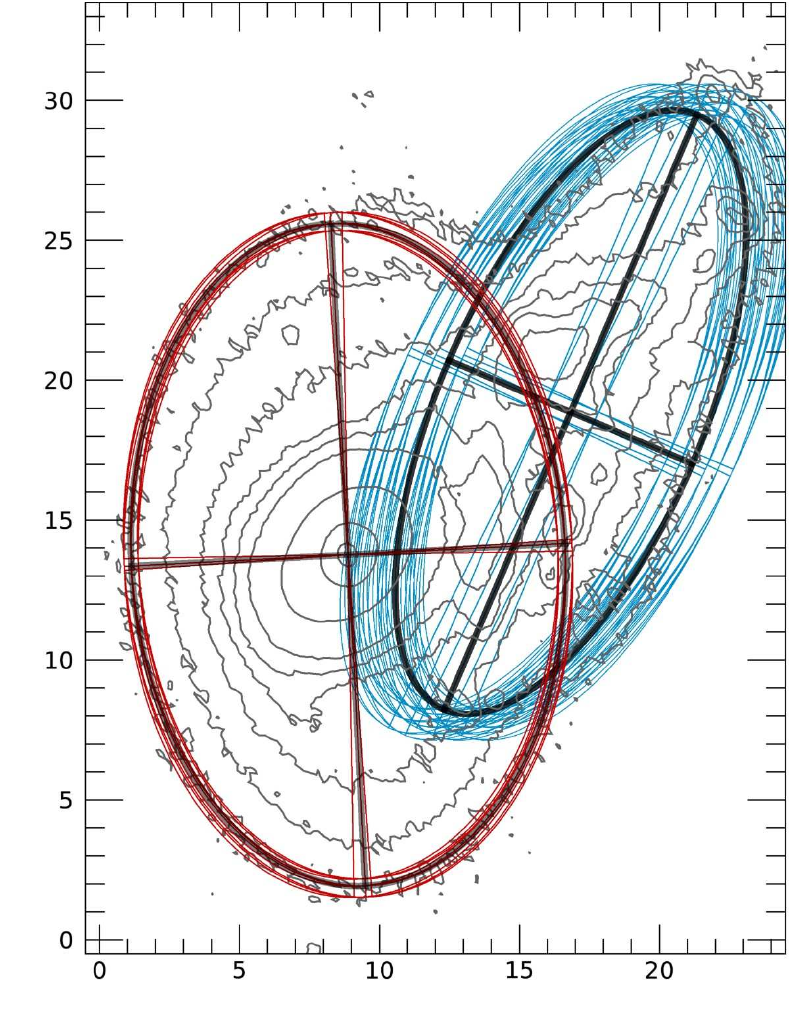}
\end{centering}
\caption{The ellipses (with relative main axes) representing the
  position, ellipticity, and orientation of the disks of the two
  galaxies are plotted in black over the $r$-band isophotes of the
  ShaSS\,073-622 system, in the same way as in Fig 19. The ellipses
  corresponding to the variants of these parameters adopted for the
  present analysis are shown in blue for ShaSS\,622 and red for
  ShaSS\,073. These values are used to estimate the uncertainties on
  the three-dimensional structure of the system, as explained in the
  text.}
\label{fig:sim}
\end{figure}




\newpage
\bibliographystyle{aasjournal}
\bibliography{biblio_GB}{}


\end{document}